\documentclass[%
 reprint,
superscriptaddress,
nofootinbib,
 amsmath,amssymb,
 aps,
onecolumn
]{revtex4-2}

\usepackage{dsfont}
\usepackage{physics}
\usepackage{mathrsfs}
\usepackage{amsmath}
\usepackage{indentfirst}
\usepackage{amssymb}
\usepackage{color}
\usepackage{amsfonts}
\usepackage[framemethod=TikZ]{mdframed}
\usepackage{mathtools}
\usepackage[retainorgcmds]{IEEEtrantools}
\usepackage[colorlinks=true,linkcolor=blue,urlcolor=blue,citecolor=blue,pdfusetitle]{hyperref}
\usepackage{times,txfonts}
\usepackage{graphicx}
\usepackage{dcolumn}
\usepackage{bm}

\usepackage[colorlinks=true,linkcolor=blue,urlcolor=blue,citecolor=blue,pdfusetitle]{hyperref}
\usepackage{hyperref}

\begin{document}

\title{Quantum switch as a thermodynamic resource in the context of passive
states}
\author{Otavio A. D. Molitor}
\email{otavio.dantasmolitor@phdstud.ug.edu.pl}
\affiliation{International Centre for Theory of Quantum Technologies (ICTQT), University of Gda\'{n}sk, 80-308 Gda\'{n}sk, Poland}
\author{\L ukasz Rudnicki}
\email{lukasz.rudnicki@ug.edu.pl}
\affiliation{International Centre for Theory of Quantum Technologies (ICTQT), University of Gda\'{n}sk, 80-308 Gda\'{n}sk, Poland}
\date{\today}

\begin{abstract}
    In recent years many works have explored possible advantages of \emph{indefinite causal order}, with the main focus on its controlled implementation known as \emph{quantum switch}. In this paper, we tackle advantages in quantum thermodynamics, studying whether quantum switch is capable of activating a passive state: either alone or with extra resources (active control state) and/or operations (measurement of the control system). By disproving the first possibility and confirming the second one, we show that quantum switch is not a thermodynamic resource in the discussed context, though, it can facilitate work extraction given external resources. We discuss our findings by considering specific examples: a qubit system subject to rotations around the $x$ and $y$ axes in the Bloch sphere, as well as general unitaries from the $U(2)$ group; and the system as a quantum harmonic oscillator with displacement operators, and with a combination of displacement and squeeze operators. 
\end{abstract}

\maketitle

\section{Introduction}

A possibility that a superposition principle can be applied to quantum operations, leading to so called \emph{indefinite causal order}, was for the first time considered in 1990 \cite{Aharonov1990}. The topic received a visible boost of attention two decades later, when Chiribella \emph{et al.} \cite{Chiribella2013} and Oreshkov \emph{et al.} \cite{Oreshkov2012} introduced the concepts of non-classical causal structures and process matrices. In this case, however, they took off from a previous work by Hardy \cite{Hardy2005}, which considers dynamical and indefinite causal structure in a potential theory of quantum gravity. A recent review \cite{Wechs2021} comprehensively covers deep theory background behind indefinite causal order treated as a quantum resource. 

The quantum switch (QS) \cite{Chiribella2013} is the paramount toy model for considerations involving indefinite causal order. It implements the controlled superposition of orders in which two (or more) unitaries (or more generally, quantum channels) are applied to a target system. 
Among others, it has been shown that with the help of QS one can get computational advantages \cite{Chiribella2013, Araujo2014, Quintino2019, Renner2022}, communication advantages \cite{Guerin2016, Ebler2018, Wei2019} or even super Heisenberg limit in metrology \cite{Zhao2020, ChapeauBlondeau2021b, Xie2021, ChapeauBlondeau2022, Yin2023}. For example, it has been predicted \cite{Ebler2018} that completely depolarizing quantum channels (which have zero capacity, i.e. they do not transmit any information), if superposed with the help of the QS, can be used for information transmition. Moreover, such an ``acausal'' superposition of noisy channels would behave as a
perfect channel \cite{Chiribella2021}, providing a possibility to reduce noise completely. Intriguingly, predicted enhancements are indeed due to superposition of orders in time \cite{Chiribella2021} and do not seem to occur fully for superposition of paths in space \cite{Abbott2020}. Interestingly, these effects seem to be more related to the sole resourcefulness \cite{Guerin2019, Jia2019, Liu2023} of indefinite causal order, rather than to particular arrangements of the involved channels. Moreover, as already mentioned, a first proposal for metrology assisted by indefinite causal order implemented through the QS has also recently been announced. In \cite{Zhao2020} the problem of estimating the product between the average position and momentum displacements has been investigated. While a basic parallel scheme with measurements of individual displacements results in an error of estimation of the product compatible with the standard quantum (shot noise) limit, a direct measurement of the two average displacements (sequential layout) allows for a quadratic improvement of the scaling -- the famous Heisenberg limit. However, the QS-assisted measurement protocol, which uses the fact that the parameter in question can be encoded in the commutator of the displacements, leads to a further quadratic improvement, the super-Heisenberg limit. Note that in this case the Weyl commutation relation between position and momentum unitary displacement operators is crucial, therefore, it is not just a bare resourcefulness of the quantum switch playing the role. The topic of metrology using indefinite causal order has further been investigated in a comprehensive way \cite{ChapeauBlondeau2021, ChapeauBlondeau2021b, ChapeauBlondeau2022, Liu2023}.

For completeness let us also briefly report experimental effort directed towards indefinite causal order, the topic in which a few proof-of-concept demonstrations have been performed to date. The first experiment with indefinite causal order implemented through QS was realised in a quantum optics context \cite{Procopio2015}. In this work a superposition of gate orders was created by considering additional degrees of freedom of photons to encode the involved qubits, with an auxiliary qubit responsible for controlling the order in which two given gates are applied to the qubit of interest. The sole concept of causality was the subject of a different experiment \cite{Rubino2017}, where an object called a causal witness, postulated in \cite{Araujo2015}, has been used to prove the \textquotedblleft acausality\textquotedblright{} of the process based on QS. Such causal witnesses serve a similar purpose to entanglement witnesses. Moreover, the quantum
switch was recently demonstrated in other experiments \cite{Goswami2018, Goswami2020, Yin2023}. Experiments with indefinite causal order are now entering the phase in which certain theoretical proposals mentioned above can be implemented: see \cite{Wei2019} for communication complexity; \cite{Guo2020,Rubino2021} for enhancements in quantum communication; \cite{Taddei2021} for computational advantages. Again on theory side, we can even observe an engineering perspective (quantum internet \cite{Caleffi2020}) entering the scene.

In the above brief review we often interchangeably treat indefinite causal order and the quantum switch. This is mainly because virtually all theoretical proposals and experimental demonstrations have to do, or even are fully concerned with QS. However, indefinite causal order is more than that, since QS does not even violate causal inequalities \cite{Araujo2015}, the latter being a true signature of acausality. On the other hand, as the metrology application shows, the emergence of extraordinary improvements might need an additional ingredient (Weyl commutation relations in this particular case), so it is not granted that indefinite causal order (or just QS) is the source of quantum advantages. These observations become particularly relevant in the context of quantum thermodynamics, where first attempts to use QS have just appeared \cite{Felce2020,Guha2020,Simonov2022,Dieguez2023}. Therefore, in the current contribution we scrutinize resourcefulness of QS in light of activation of passive states in quantum thermodynamics. While it was initially treated in \cite{Simonov2022}, here the activation of passive states by means of the QS is further explored and new situations are explored. First, we study the activation with and without measuring and tracing out the control degree-of-freedom, in which the latter depends on non-diagonal elements in the Hamiltonian of the control. Second, in our case at all times we treat the state of the control and its measurement basis in the Bloch sphere representation, which experimentally speaking might have more impact. Third, the conditions for state activation are derived by looking at energy differences before and after applying the QS (and post-selection). Hence, whenever this energy difference is negative, one certifies state activation. Finally, when it comes to concrete examples of systems, we show that for two-level systems (qubit) depending on the chosen measurement basis no work can be extracted, even if the control was initially prepared with quantum coherence. Moreover, we also consider a quantum harmonic oscillator as the system, which has a whole set of peculiarities when compared to finite-level systems (e.g. we can continuously set the unitaries to the identity operator, which might lead to singularities in the renormalization after measurement of the control). The case of applying the QS to systems with infinite levels -- to which the quantum harmonic oscillator is an example -- was to our best knowledge just treated previously in \cite{Giacomini2016}, which was done in a very abstract manner using the process matrix formalism. In this paper we go to a more concrete scenario, using well-known examples of unitaries for continuous-variable systems (displacement and squeeze operators\footnote{This name is not consistent in the literature, as it is as well called ``squeezing operator'' by some authors. In this paper we stick to the term ``squeeze operator''.}). Therefore, since electromagnetic modes can be treated as quantum harmonic oscillators, the study here provided can have an impact in quantum optic setups.

This paper is organized as follows. In Sec. \ref{sec:qsandthermo} we expand the discussion concerned with the interplay between quantum thermodynamics and the quantum switch. We observe that, while QS can be used in a clever way to activate \cite{Simonov2022}, so called, passive states \cite{Alicki2018}, the question about an origin of necessary resources is, due to a special status of thermodynamics, perhaps more relevant than in other scenarios mentioned above. Therefore, we pose a very precise question pertaining to the problem of passive states subject to QS, in order to figure out whether thermodynamic advantages in this context can come from the sole resourcefulness of QS, or they rather come from the ancillary degrees of freedom. In Sec. \ref{sec:results} we show that the latter scenario applies to the setup under consideration. In order to see how the results work in different cases, we apply the framework to specific examples in Sec. \ref{sec:examples}. First, in Sec. \ref{sec:twolevel} the situation considered is when the system is a qubit (only two-levels) for different unitaries: (i) when they are rotations around the $x$ and $y$ axes in the Bloch sphere (Sec. \ref{sec:rotation}) and (ii) when they are represented as general unitaries from the $U(2)$ group (Sec. \ref{sec:generalU2}). Then, we study the case in which the system is a quantum harmonic oscillator (Sec. \ref{sec:qho}) with two different combinations of unitaries: (i) both unitaries are displacement operators (Sec. \ref{sec:displacement}) and (ii) one unitary is a displacement operator and the second is a squeeze operator (Sec. \ref{sec:displacementsqueeze}). Finally, we pass to Sec. \ref{sec:conclusions}, where we draw conclusions about this work.

\section{Quantum switch and passive states in thermodynamics}\label{sec:qsandthermo}

Whenever QS has been shown to provide quantum advantages, a similar scheme is being exercised. One considers a state of the system, denoted by $\rho_{S}$, and a control qubit, denoted by $\rho_{C}$. Initially, the state of the total system is $\rho_{SC}=\rho_{S}\otimes\rho_{C}$, so by assumption there is no correlation between the system and the control qubit at the initial time.

Given two unitaries $U_{1}$ and $U_{2}$ (or other quantum channels, as everything naturally extends to Kraus decomposition involving more terms, see \cite{Ebler2018} as a profound example), one performs the operation  
\begin{equation}
\label{eq:qsunitary}
U_{\textrm{QS}}=U_{2}U_{1} \otimes \ketbra{0}{0}_C+U_{1}U_{2}\otimes\ketbra{1}{1}_C.
\end{equation}
Note that, from a physical point of view, in the above definition we give meaning to the computational basis of the control qubit, $\{\ket{0} , \ket{1}\} $. Afterwards, one measures the control qubit in a \emph{suitable} basis (most often in $\{\ket{+} , \ket{-}\} $) and infers conclusions about gains associated with the system. 

Let us now consider the problem of passive states in thermodynamics. The state $\rho_{S}$ is passive with respect to the system Hamiltonian $H_{S}$, if \cite{Haag1974, Lenard1978, Pusz1978}
\begin{equation}
\tr\left\{\rho_{S}H_{S}\right\} \leq \tr\left\{U\rho_{S}U^{\dagger}H_{S}\right\},
\end{equation}
for every unitary operation $U$. The notion closely related with passivity of states is ergotropy \cite{Allahverdyan2004, Safranek2023} as it measures extractable work. The definition of ergotropy is the following: consider a quantum system whose state is $\rho_S$ and Hamiltonian $H_S$, the ergotropy of such system -- or \emph{extractable work} -- is expressed by
\begin{equation}
    W_\text{max} := \tr \{ \rho_S H_S \} - \underset{U}{\text{min}} \tr \{ U \rho_S U^\dagger H_S \}
\end{equation}
where the minimization procedure is applied to all the unitary transformations existing in the Hilbert space of the system $\mathcal{H}_S$. As shown in \cite{Alicki2013}, the ergotropy is upper bounded as the following:
\begin{equation}
    W_\text{max} \leq \tr \{ \rho_S H_S \} - \tr \{ \sigma_{\beta} H_S \}
\end{equation}
with 
\begin{equation}
    \sigma_{\beta} = \frac{e^{-\beta H_S}}{Z_S}
\end{equation}
being the Gibbs thermal state at inverse temperature $\beta$, such that its von Neumann entropy $S(\rho) := -\tr \{ \rho \ln \rho \}$ is the same as of $\rho_S$, i.e. $S(\rho_S) = S(\sigma_{\beta})$. Moreover, $Z_S$ is the partition function of the system, defined by $Z_S := \tr \{ \text{exp} (-\beta H_S) \}$. Therefore, it can be seen that the Gibbs thermal state sets a limit on the amount of work that can be extracted from a quantum system. As a matter of fact, having a system in a thermal state means that no work can be extracted from it using any unitary $U$.

One then is led to a question: is it possible to extract work from passive states with the help of QS? The natural way is to check whether starting from a passive state one can obtain a state whose ergotropy is non-zero. In this line, Simonov \emph{et al.} \cite{Simonov2022} scrutinized  gains in ergotropy due to application of QS. As usual, it turded out that occurrence of the potential benefits critically depends on the basis in which
the control qubit is being measured. This fact suggests the follow-up question. Is the increased ergotropy a consequence of extra information (like in the Maxwell demon problem) which is available after tailored
measurements applied to the control qubit (i.e. some resources associated with the control qubit), or is it rather the acausal character of the quantum switch which plays the major role? Here, we are going to discuss the second possibility.

To this end we resort to the fact, that the notion of passivity admits the phenomenon of \emph{superadditivity}. While two states can individually
be passive, its tensor product does not need to be such. Only Gibbs states are completely passive, which means they do not admit superadditivity in that context.

\section{Results}
\label{sec:results}

We are in position to formalize our problem at hand. As before, we assume that at an initial time the total state is not correlated $\rho_{SC}=\rho_{S}\otimes\rho_{C}$, and moreover, both $\rho_{S}$ and $\rho_{C}$ are individually passive with respect to their local Hamiltonians $H_{S}$ and $H_{C}$. Clearly, due to the phenomenon of superadditivity, if we admit any unitaries acting on the composite system, we will
potentially be able to extract work from $\rho_{SC}$, as the composite state is not necessarily passive. 

We however restrict the set of allowed global unitary operations to such which are realized by QS involving two unitaries on the system side only. We are then to check whether this setting is enough to activate
the system, i.e. to observe
\begin{eqnarray}
\Delta_\text{QS} &:=& \tr\left\{U_{\text{QS}}\rho_{SC} U_{\text{QS}}^{\dagger}H_{SC}\right\}-\tr\left\{\rho_{SC} H_{SC}\right\} \nonumber\\
&=& E_{SC}' - E_{SC} < 0,
\end{eqnarray}
with the total Hamiltonian $H_{SC}=H_{S}\otimes\mathds{1}_{C}+\mathds{1}_{S}\otimes H_{C}$. Here we consider that system and control do not interact, hence $E_{SC} = E_S + E_C$. 

Since application of ``causally separable'' unitaries on the system side is not enough for activation to occur, what we shall do is to check whether including QS is already a sufficient resource for the discussed purpose. To this end we can explicitly compute
\begin{equation}
E_{SC}' = \bra{0}\rho_{C}\ket{0} E_{12} + \bra{1}  \rho_{C}  \ket{1} E_{21} + \bra{0}\! \rho_{C}  \ketbra{0}{0}  H_{C}  \ket{0} + \bra{1}  \rho_{C}  \ketbra{1}{1}  H_{C} \ket{1} +\chi\bra{0}  \rho_{C}  \ketbra{1}{1}  H_{C} \ket{0} + \chi^*\bra{1}  \rho_{C}  \ketbra{0}{0}  H_{C}\ket{1},
\end{equation}
with
\begin{equation}
    E_{12} := \tr \left \{ U_{2}U_{1}\rho_{S}U_{1}^{\dagger}U_{2}^{\dagger}H_{S}\right \}, \qquad E_{21} := \tr \left\{U_{1}U_{2}\rho_{S}U_{2}^{\dagger}U_{1}^{\dagger}H_{S}\right\},
\end{equation}
and
\begin{equation}
\label{eq:chi}
\chi=\tr\left\{U_{2}U_{1}\rho_{S}U_{2}^{\dagger}U_{1}^{\dagger}\right\}\equiv\left|\chi\right|e^{i\phi}.
\end{equation}
The complex number $\chi$ is connected to the unitary cross-map~\cite{Simonov2022}, encoding correlations relevant when both unitaries do not commute. In the communing case we trivially get $\chi=1$. Let us denote:
\begin{equation}
U_{\phi,\pm}=\left(\begin{array}{cc}
1 & 0\\
0 & \pm e^{-i\phi}
\end{array}\right),
\end{equation}
and introduce states
\begin{equation}
\tilde{\rho}_{C}=\frac{1+\left|\chi\right|}{2}U_{\phi,+}\rho_{C}U_{\phi,+}^{\dagger}+\frac{1-\left|\chi\right|}{2}U_{\phi,-}\rho_{C}U_{\phi,-}^{\dagger},
\end{equation}
\begin{equation}
\tilde{\rho}_{S}=\bra{0}\rho_{C}\ket{0} U_{2}U_{1}\rho_{S}U_{1}^{\dagger}U_{2}^{\dagger}+\bra{1}\rho_{C}\ket{1} U_{1}U_{2}\rho_{S}U_{2}^{\dagger}U_{1}^{\dagger}.
\end{equation}
We find that
\begin{equation}
E_{SC}' = \tilde{E}_S + \tilde{E}_C.
\end{equation}
where by analogy we define
\begin{equation}
    \tilde{E}_S := \tr\{ \tilde{\rho}_S H_S \}, \qquad \tilde{E}_C := \tr\{ \tilde{\rho}_C H_C \}.
\end{equation}
Since both $\tilde{\rho}_{S}$ and $\tilde{\rho}_{C}$ represent the result of applying incoherent (convex) superpositions of local unitary operations to $\rho_{S}$ and $\rho_{C}$ respectively, we conclude that
\begin{equation}
E_S + E_C \leq \tilde{E}_S + \tilde{E}_C.
\end{equation}
Consequently
\begin{equation}
\Delta_\text{QS} \geq 0
\end{equation}
since from individual passivity of $\tilde{\rho}_{S}$ and $\tilde{\rho}_{C}$
we know that
\begin{equation}
\label{eq:deltaS}
\Delta_S := \tilde{E}_S - E_S \geq 0, \qquad \Delta_C := \tilde{E}_C - E_C \geq 0.
\end{equation}

As our first result, we find that the QS itself is incapable of performing an activation  of a passive state. While it has been expected that the passive control state will not become activated (no unitaries are operating on these degrees of freedom), the same is shown to be true for the reduced system state itself, and as a mere consequence of linearity extends to the composite state of system and control. Therefore, the predicted increase of the ergotropy of the composite system seems to be associated with resources of the control qubit and/or measurements performed on it, rather than the sole action of the quantum switch. 

To frame the first possibility we suppose that the control qubit is in a generic state that does not need to be passive. As a direct calculation shows
\begin{equation}
    \Delta_C = 2 \Re\{ \bra{0}\rho_C \ketbra{1}{1} H_C \ket{0} (\chi - 1) \},
\end{equation}
where $\Re\{z\}$ denotes the real part of the complex number $z$, 
so it is straightforward to minimize this expression with respect to the state of the qubit. To this end one needs $\bra{0}\rho_C \ket{1}=-e^{i \phi}/\sqrt{2}$, where the phase $\phi$ is selected to cancel the phase of $\bra{1} H_C \ket{0} (\chi - 1)$. Consequently
\begin{equation}
    \min_{\rho_C}\Delta_\text{C} = -\sqrt{2} \left|\bra{1} H_C \ket{0} (\chi - 1)\right|.
\end{equation}
The last expression does not only show an expected effect of activation for the control, but also proves that the composite system plus control state activation is possible if and only if  $\chi \neq 1$ and  the Hamiltonian $H_C$ has non-diagonal terms in the computational basis $\{\ket{0}, \ket{1} \}$ defined by the action of the QS. 
Non-commuting unitaries $U_1$ and $U_2$ are essential for state activation, otherwise $\chi = 1$. To get $\Delta_\text{QS} \leq 0$ one needs a sufficiently large value of the control Hamiltonian coherence in comparison with the energy scale of the system. We stress that in this way it is impossible to activate just the system. The activation can occur only for the composite state.

On the other hand, even if the requirement of Hamiltonian of the control having non-diagonal terms in the computational basis is not fulfilled, it is still possible to activate the state of the system by measuring the control qubit. In order to show how it happens, consider that the state of the control is pure $\rho_C = \ketbra{\psi}{\psi}_C$ and parametrized in the following way:
\begin{equation}
    \ket{\psi}_C = \cos(\frac{\theta_C}{2}) \ket{0}_C + e^{i \varphi_C}\sin(\frac{\theta_C}{2}) \ket{1}_C
\end{equation}
such that in the Bloch sphere it is represented by the point $(\sin \theta_C \cos\varphi_C, \sin\theta_C \sin\varphi_C, \cos\theta_C)$. Then, the final joint state of system and control is equal to
\begin{equation}
    \begin{aligned}
        \rho_{SC}' &= \cos^2 \left ( \frac{\theta_C}{2} \right ) U_2 U_1 \rho_S U_1^\dagger U_2^\dagger \otimes \ketbra{0}{0}_C 
        + \frac{e^{i \varphi_C}}{2} \sin\theta_C\, U_2 U_1 \rho_S U_2^\dagger U_1^\dagger \otimes \ketbra{0}{1}_C\\
        &+ \frac{e^{-i \varphi_C}}{2} \sin\theta_C\, U_1 U_2 \rho_S U_1^\dagger U_2^\dagger \otimes \ketbra{1}{0}_C
        + \sin^2\left ( \frac{\theta_C}{2} \right ) U_1 U_2 \rho_S U_2^\dagger U_1^\dagger \otimes \ketbra{1}{1}_C.
    \end{aligned}
\end{equation}
The measurement of the control is supposed to be done by means of a projective measurement onto the state
\begin{equation}
    \ket{\psi_M}_C = \cos(\frac{\theta_M}{2}) \ket{0}_C + e^{i \varphi_M}\sin(\frac{\theta_M}{2}) \ket{1}_C
\end{equation}
so that the state of the system post-measurement of the control is
\begin{equation}
    \rho_{S,M} = \frac{(\mathds{1}_S \otimes \bra{\psi_M}_C)\rho_{SC}'(\mathds{1}_S \otimes \ket{\psi_M}_C)}{\tr \{ (\mathds{1}_S \otimes \bra{\psi_M}_C)\rho_{SC}'(\mathds{1}_S \otimes \ket{\psi_M}_C) \}}.
\end{equation}
This state explicitly reads
\begin{equation}
\label{eq:statepostmeas}
    \begin{aligned}
        \rho_{S,M} &= \frac{1}{N_M} \bigg ( \cos^2\left( \frac{\theta_C}{2} \right) \cos^2\left( \frac{\theta_M}{2} \right) \, U_2 U_1 \rho_S U_1^\dagger U_2^\dagger
        + \sin^2\left( \frac{\theta_C}{2} \right) \sin^2\left( \frac{\theta_M}{2} \right) \, U_1 U_2 \rho_S U_2^\dagger U_1^\dagger \\
        &+ \frac{e^{-i (\varphi_C + \varphi_M)}}{4} \sin\theta_M \sin\theta_C\, U_1 U_2 \rho_S U_1^\dagger U_2^\dagger 
        + \frac{e^{i (\varphi_C + \varphi_M)}}{4} \sin\theta_M \sin\theta_C\, U_2 U_1 \rho_S U_2^\dagger U_1^\dagger \bigg )
    \end{aligned}
\end{equation}
with
\begin{equation}
    N_M = \frac{1}{2}(1 + \cos\theta_C \cos\theta_M + \sin\theta_C \sin\theta_M\Re \{ \chi e^{i(\varphi_C + \varphi_M)})  \}).
\end{equation}

As one can see, the state expressed by Eq. (\ref{eq:statepostmeas}) contains not only the incoherent terms (the ones associated with the diagonal elements of $\rho_C$), but also coherences coming from the off-diagonal terms of the state of the control. Hence, the average internal energy of the system post-application of the QS and post-measurement of the control,
\begin{equation}
    E_{S,M} := \tr \{ \rho_{S,M} H_S \} 
\end{equation}
might be inferior than the initial average internal energy. As in the pre-measurement case, coherences in the control qubit are mandatory for state activation. In fact, one can see that whenever the difference between the final ($E_{S,M}$) and initial value is negative, that is
\begin{equation}
    \Delta_{S,M} := E_{S,M} - E_S < 0
\end{equation}
there is activation of the state of the system, and work can be extracted from it. 

By expanding the previous expression for $\Delta_{S,M}$, one finds that it is equivalent to:
\begin{equation}
        \Delta_{S,M} = \frac{1}{N_M} \bigg ( \cos^2\left(\frac{\theta_C}{2}\right)\cos^2\left(\frac{\theta_M}{2}\right) \Delta_{12} + \sin^2\left(\frac{\theta_C}{2}\right)\sin^2\left(\frac{\theta_M}{2}\right) \Delta_{21} + \frac{1}{2} \sin\theta_C \sin\theta_M \Re \{ \Delta_F e^{i(\varphi_C + \varphi_M)} \} \bigg )
\end{equation}
where
\begin{equation}
    \Delta_{12} := \tr \{ U_2 U_1 \rho_S U_1^\dagger U_2^\dagger \} - E_S, \qquad  \Delta_{21} := \tr \{ U_1 U_2 \rho_S U_2^\dagger U_1^\dagger \} - E_S
\end{equation}
and 
\begin{equation}
    \Delta_F := F_S - \chi E_S,\qquad F_S := \tr \{U_2 U_1 \rho_S U_2^\dagger U_1^\dagger H_S\}.
\end{equation}
Since $\Delta_{12} \geq 0$ and $\Delta_{21} \geq 0$, one finds the conditions for a possible state activation after measuring the control: (i) $\theta_C \neq 0, \pi$ and $\theta_M \neq 0, \pi$ (i.e. the states cannot be either $\ket{0}$ or $\ket{1}$); (ii) $\tan(\varphi_C + \varphi_M) \neq \Re \{ \Delta_F \}/\Im \{ \Delta_F \}$, and (iii) $\sin\theta_C\sin\theta_M\Re\{\Delta_F e^{i(\varphi_C + \varphi_M)}\} < 0$. These are necessary, but not sufficient conditions for state activation of the system.

\begin{figure}[h!]
    \centering
    \includegraphics[width=0.43\textwidth]{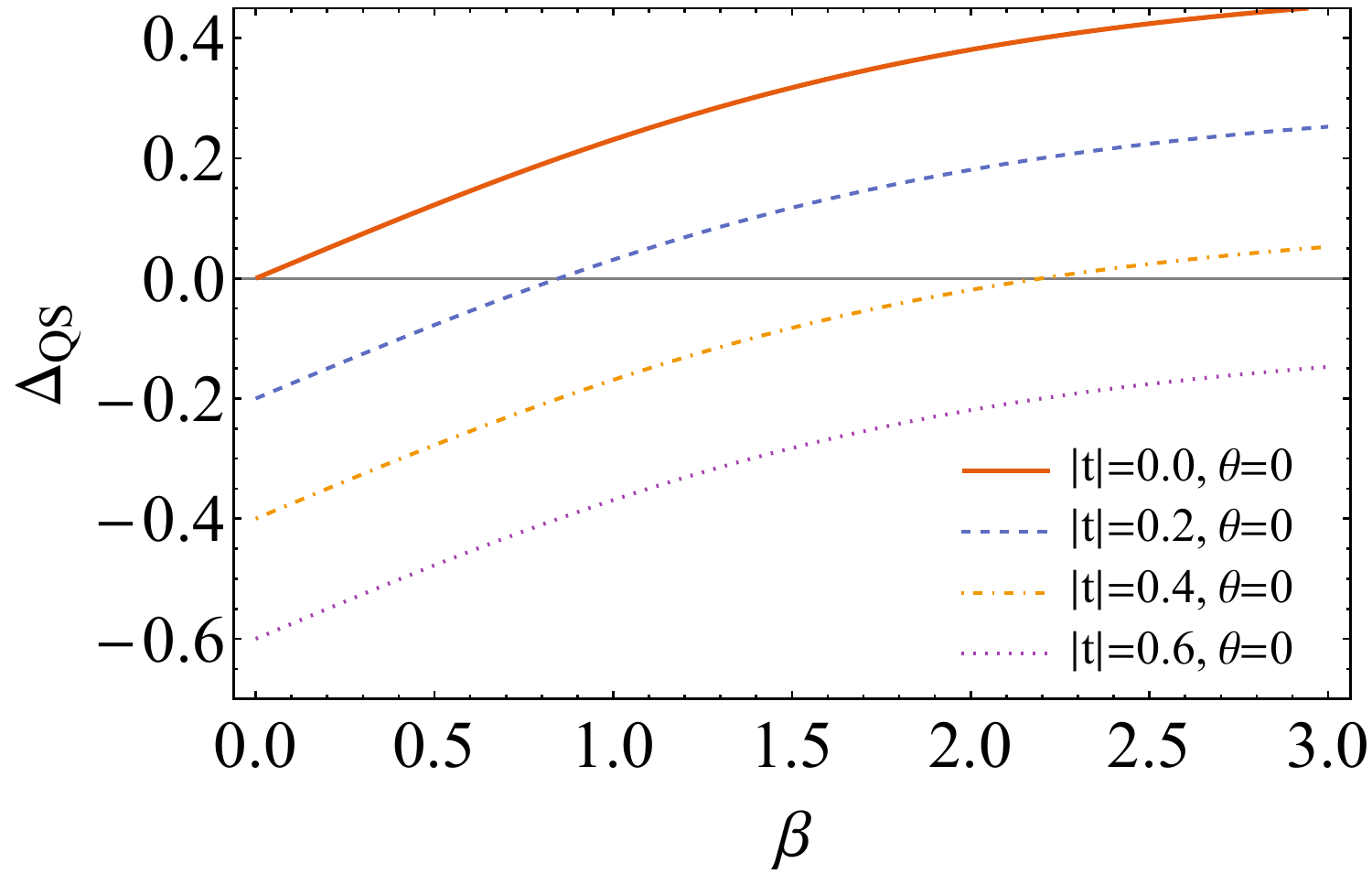}
    \caption{Plot of $\Delta_\text{QS}$ as a function of $\beta$, for fixed $\omega=1.0$, $\alpha_x=\pi/2$, $\alpha_y=\pi$ and different values of $t$, when the unitaries are rotations around the $x$ and $y$ axes. As it is evidenced by the plots, for $|t|=0$ no activation occurs ($\Delta_\text{QS} \geq 0$). However, as $|t|$ increases, more energy can be extracted from system plus control, and for a larger range of inverse temperatures $\beta$.}
    \label{fig:deltaqsbetat}
\end{figure}

\section{Examples}
\label{sec:examples}

Let us now test the above considerations with concrete scenarios. We start with the situation in which the system has only two-levels (qubit) and the unitaries are rotations around the $x$ and $y$ axes of the Bloch sphere (Sec. \ref{sec:rotation}). Still in the qubit scenario, we then consider general $U(2)$ unitaries in Sec. \ref{sec:generalU2}. In the continuation, we pass to the case in which the system consists of a quantum harmonic oscillator in two different combinations of unitaries: (i) both being displacement operators (Sec. \ref{sec:displacement}) and (ii) one unitary being the displacement operator and the other the squeeze operator (Sec. \ref{sec:displacementsqueeze}). In all parts, units are such that $\hbar = k_B = 1$.

\subsection{Two-level systems}
\label{sec:twolevel}

Consider the case in which both system and control are represented by two-level systems (qubits). In such a case, the individual Hamiltonian of the system is written as 
\begin{equation}
    H_S = \frac{\omega}{2} (\mathds{1}_S - \sigma^z_S)
\end{equation}
where $\mathds{1}_S$ is the identity operator living in the Hilbert space of the system , and $\sigma^z_S$ is the ``z'' Pauli matrix in $\mathcal{H}_S$. The Hamiltonian of the control is similar, but it contains a non-diagonal term,
\begin{equation}
    H_C = \frac{\omega}{2} (\mathds{1}_C - \sigma^z_C) + t \ketbra{0}{1}_C + t^* \ketbra{1}{0}_C
\end{equation}
with the same identity and Pauli operators as before, but now living in $\mathcal{H}_C$ and $t = |t|e^{i\theta}\in \mathbb{C}$ is connected to the probability that the control qubit will jump from one state to the other. For simplicity, we consider that system and control are resonant ($\omega_S = \omega_C = \omega$) and the total Hamiltonian is a non-interacting one $H_{SC} = H_S \otimes \mathds{1}_C + \mathds{1}_S \otimes H_C$. The initial state of the system is taken to be the Gibbs state,
\begin{equation}
    \rho_S = \frac{e^{-\beta H_S}}{Z_S} = 
    \begin{pmatrix}
        \frac{1}{1 + e^{-\beta \omega}} && 0 \\
        0 && 1-\frac{1}{1 + e^{-\beta \omega}}
    \end{pmatrix}
\end{equation}
with $Z_S = \tr\{ e^{-\beta H_S} \}$ and $\beta$ is the inverse temperature of the system. Moreover, the control is initially prepared in a generic pure state $\rho_C = \ketbra{\psi}{\psi}_C$, with $\ket{\psi}_C = \cos(\theta_C/2)\ket{0}_C + e^{i \varphi_C}\sin(\theta_C/2)\ket{1}_C$, and $\theta_C\in [0,\pi]$, $\varphi_C \in [0,2\pi]$. The initial joint state is a non-correlated, product state $\rho_{SC} = \rho_S \otimes \rho_C$. We then consider two different scenarios for the unitaries: (i) they correspond to rotations around the $x$ and $y$ axes in the Bloch sphere and (ii) general $U(2)$ unitaries (which themselves are decomposed as rotations in the Bloch sphere).
\begin{figure}[h!]
    \centering
    \includegraphics[width=0.7\textwidth]{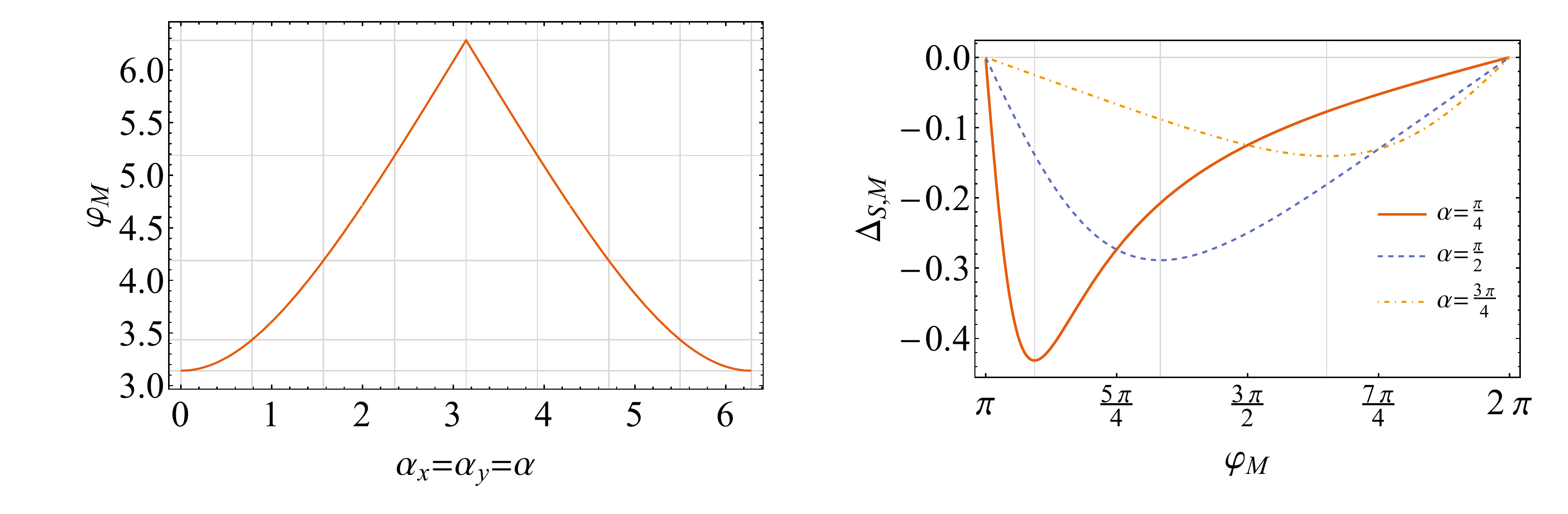}
    \caption{Plots of the value of $\varphi_M$ (in radians) for which $\Delta_{S,M}$ is minimum (left) and of $\Delta_{S,M}$ as a function of $\varphi_M$ (in radians) for different $\alpha$ (right). In both we consider the state of the control to be equal to $\ketbra{+}{+}_C$ ($\theta_C=\pi/2$, $\varphi_C=0$), $\omega=1.0$, $\theta_M=\pi/2$ and $\beta\to0$.}
    \label{fig:maxDeltaSMvarphiM}
\end{figure}

\subsubsection{Rotation operators}
\label{sec:rotation}

First, we start with unitaries as rotations around the $x$ and $y$ axes in the Bloch sphere,
\begin{equation}
    U_1 = R_x(\alpha_x) = e^{-i \sigma_x \alpha_x/2}, \qquad  U_2 = R_y(\alpha_y) = e^{-i \sigma_y \alpha_y/2}
\end{equation}
where $\sigma_x$ and $\sigma_y$ are Pauli matrices, and $\alpha_x, \alpha_y \in [0,2\pi]$ are the angles of rotation. The final state after applying the QS is then denoted $\rho_{SC}' = U_\text{QS}\rho_{SC}U_\text{QS}^\dagger$, with $U_\text{QS}$ as in Eq. (\ref{eq:qsunitary}). With all that it is possible to calculate
\begin{equation}
        \Delta_\text{QS} = \frac{\omega}{2} \Bigg[1 - \cos\alpha_x\cos\alpha_y + \frac{|t|}{\omega} \sin\alpha_x\sin\alpha_y\sin\theta_C\sin(\theta+\varphi_C) \Big )\tanh\left( \frac{\beta \omega}{2} \right) \Bigg] - 2|t|\cos(\theta+\varphi_C)\sin\theta_C\sin^2\left(\frac{\alpha_x}{2}\right)\sin^2\left(\frac{\alpha_y}{2}\right). 
\end{equation}
This expression is plotted in Fig.\ref{fig:deltaqsbetat} in the case that the state of the control corresponds to the pure state $\ket{+} = (\ket{0}+\ket{1})/\sqrt{2}$. As expected from the calculations in Sec. \ref{sec:results}, when the Hamiltonian of the control is diagonal in the computational basis ($|t|=0$), the energy of the system plus control is always higher than the initial energy after applying the quantum switch, for whatever inverse temperature $\beta$. As one increases the value of $|t|$, it is possible to reach lower final energy for a range of $\beta$, meaning that the state of system plus control is activated.

Now passing to the case that the control is measured by means of a projector in the state $\ket{\psi_M}_C = \sin(\theta_M/2)\ket{0}_C + e^{i\varphi_M}\sin(\theta_M/2)\ket{1}_C$, we have the following necessary (but not sufficient) conditions for state activation:
\begin{equation}
    \sin\theta_C \neq 0, \quad \sin\theta_M \neq 0
\end{equation}
\begin{equation}
    \tan(\varphi_C + \varphi_M) \neq (\cot\alpha_x \cot\alpha_y - \csc\alpha_x\csc\alpha_y)\sinh\beta\omega
\end{equation}
and 
\begin{equation}
        \frac{\omega \sin\theta_C\sin\theta_M}{2(1+e^{\beta\omega})^2} ((e^{2\beta\omega} - 1)(1-\cos\alpha_x\cos\alpha_y)\cos(\varphi_C+\varphi_M) + 2e^{\beta\omega}\sin\alpha_x\sin\alpha_y\sin(\varphi_C+\varphi_M)) < 0.
\end{equation}
These get simplified when $\beta\to 0$ (i.e. the Gibbs thermal state corresponds to the maximally mixed state $\rho_S = \mathds{1}_S/2$) and $\alpha_x=\alpha_y=\alpha$, then the last two conditions become:
\begin{equation}
    \tan(\varphi_C+\varphi_M) \neq 0
\end{equation}
and
\begin{equation}
    \sin\theta_C\sin\theta_M\sin(\varphi_C+\varphi_M) < 0
\end{equation}
then, if we set, for example, $\theta_C = \pi/2$ and $\varphi_C=0$ -- which corresponds to the state $\ketbra{+}{+}_C$ -- one has that $\varphi_M \neq 0,\pi$ and $\sin\theta_M\sin\varphi_M < 0$, meaning that one must have $\varphi_M\in\,\, ]\pi,2\pi[$ for possible state activation. 

In this simplified scenario (state of the control $\ketbra{+}{+}_C$, $\beta\to 0$ and $\alpha_x=\alpha_y=\alpha$), $\Delta_{S,M}$ becomes:
\begin{equation}
    \Delta_{S,M} = \frac{\omega \sin\varphi_M}{2\cos\varphi_M + 4\cos\varphi_M\cot\alpha\csc\alpha + 4\csc^2\alpha\csc\theta_M}
\end{equation}
which clearly has a minimum for $\theta_M = \pi/2$ (remember that $\theta_M \in \,\, ]0,\pi[\,\,$). Then we plot in Fig. \ref{fig:maxDeltaSMvarphiM} the angle $\varphi_M$ for which we obtain minimum $\Delta_{S,M}$, as well as the values of the latter as a function of $\varphi_M$ for different values of $\alpha$. Here, as a matter of fact, the previous conditions for state activation are not only necessary, but also sufficient, since when $\beta\to 0$ the ``causally ordered'' energy differences are equal to zero: $\Delta_{12}=\Delta_{21}=0$.

\begin{figure}[h!]
    \centering
    \includegraphics[width=\textwidth]{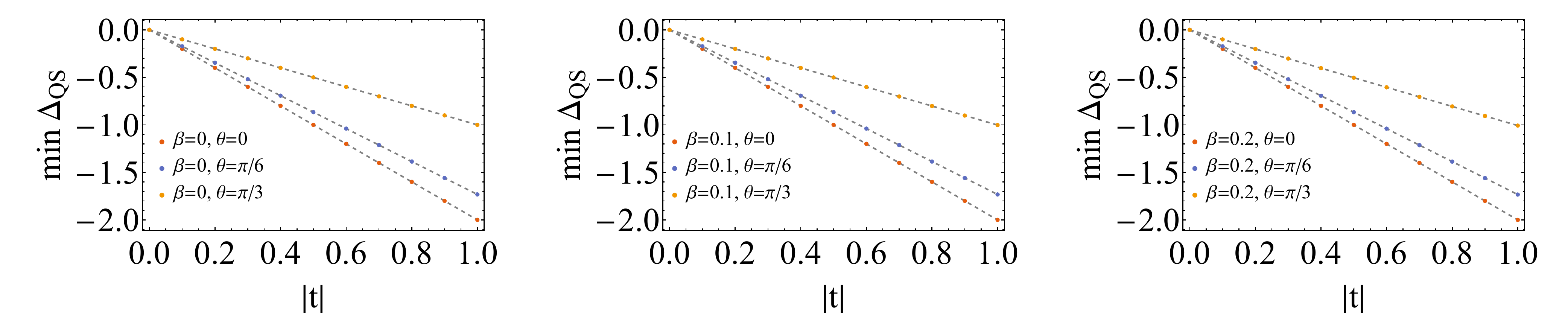}
    \caption{Minimum values achieved by $\Delta_\text{QS}$ as a function of $|t|$, for fixed $\omega=1.0$ and for different inverse temperatures and angles $\theta$ ($\beta=0$ on the left, $\beta=0.1$ on the middle and $\beta=0.2$ on the right). For each point, a different combination of $\lambda_k, \gamma_k, \delta_k$ leads to the minimum value of $\Delta_\text{QS}$. Here the state of the control is set to be the $\ketbra{+}{+}_C$ state ($\theta_C=\pi/2$, $\varphi_C=0$). Curiously, the slopes of the dashed lines from the plots do not depend on $\beta$, but solely on $\theta$.}
    \label{fig:deltaqsbetatU2}
\end{figure}

\subsubsection{General U(2) unitaries}
\label{sec:generalU2}

In the most general scenario, unitary operations in $\mathcal{L}(\mathcal{H}_2)$, with $\mathcal{H}_2$ being the Hilbert space of dimension 2 (qubits), can be written in the generic $U(2)$ representation group \cite{Nielsen2010},
\begin{equation}
    U_k = e^{i \alpha_k} R_z (\lambda_k) R_y (\gamma_k) R_z (\delta_k), \quad k = 1,2
\end{equation}
with $\alpha_k, \lambda_k, \gamma_k, \delta_k \in \mathbb{R}$ and $R_y, R_z$ are rotations around the $y$ and $z$ axis of the Bloch sphere, respectively. Unfortunately, compact expressions cannot be obtained here, but numerically we can try to find combinations of $\alpha_k, \lambda_k, \gamma_k, \delta_k$ that minimize $\Delta_\text{QS}$ for a given combination of $\omega, \beta$ and $t$ (actually $\alpha_k$ do not matter in this case), given a certain state of the control, which we take to be the $\ketbra{+}{+}_C$ state. In Fig. \ref{fig:deltaqsbetatU2}, one finds the plots of the minimum value of $\Delta_\text{QS}$ as a function of $|t|$ for a few inverse temperatures $\beta$ and angles $\theta$. We see that for each $|t|>0$ it is always possible to reach negative $\Delta_\text{QS}$ and curiously, all the points with the same angle $\theta$  converge to the same line with a well defined slope, irrespective of the inverse temperature $\beta$. Then, one might take the simplifying scenario when $\beta \to 0$:
\begin{equation}
    \Delta_\text{QS}^{\beta\to0} = - \left ( \frac{-12 + \epsilon (\lambda_1,\gamma_1,\delta_1,\lambda_2,\gamma_2,\delta_2)}{16} \right ) \cos \theta \, |t|
\end{equation}
where 
\begin{equation}
    \text{min}\,\epsilon (\lambda_1,\gamma_1,\delta_1,\lambda_2,\gamma_2,\delta_2) \equiv -20,
\end{equation}
such that the minimum value of the energy difference of system plus control is totally determined by the off-diagonal term of the Hamiltonian of the control $t$ (of course, it happens for different combinations of $\lambda_k$, $\gamma_k$ and $\delta_k$).

\begin{figure}[h!]
    \centering
    \includegraphics[width=0.4\textwidth]{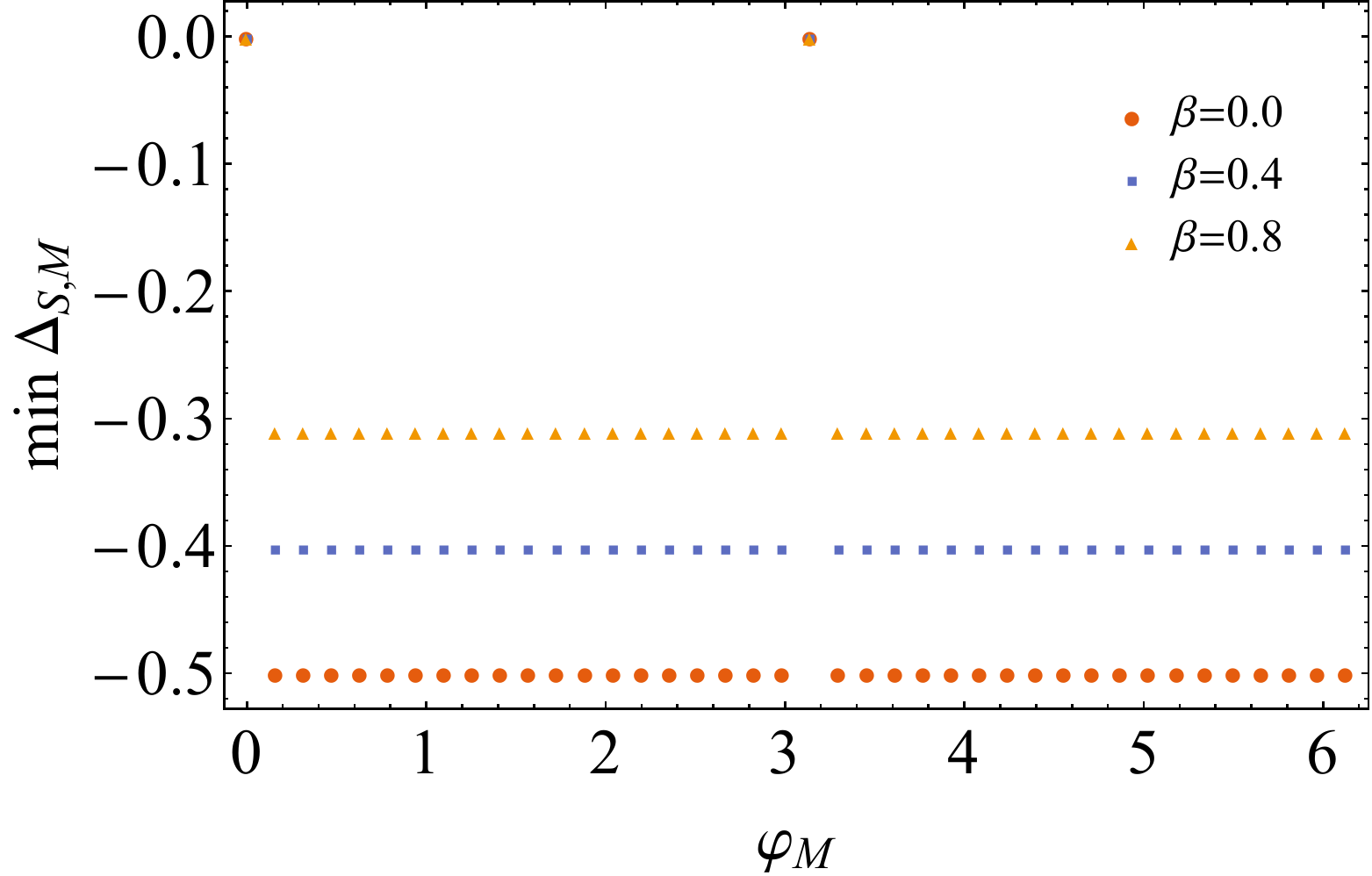}
    \caption{Minimum value achieved by $\Delta_{S,M}$ as a function of $\varphi_M$, when $\omega=1.0$, $\theta_C=\theta_M=\pi/2$ and $\varphi_C=0$, for different $\beta$. For a better visualization, we plot just a few points, in order to show that the minimum value of $\Delta_{S,M}$ is a constant, except for $\varphi_M=0,\pi$, when no state activation is possible ($\text{min} \,\Delta_{S,M}=0$).}
    \label{fig:minDeltaSMvarphiM}
\end{figure}

\begin{figure}[t]
        \centering
        \includegraphics[width=1.0\textwidth]{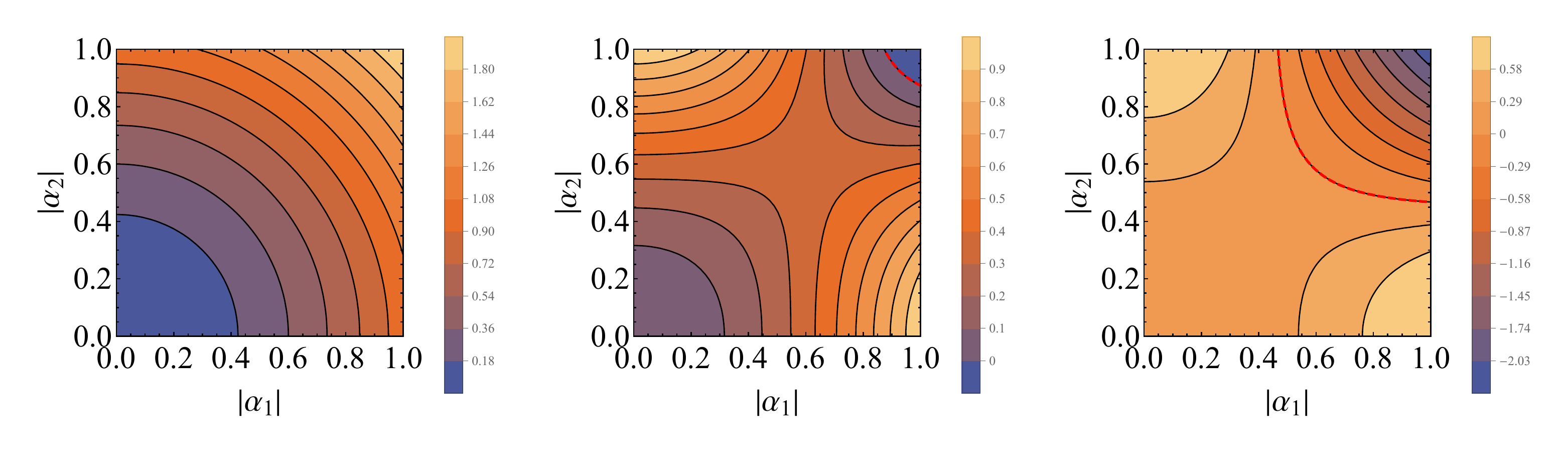}
        \caption{Plots of $\Delta_\text{QS}$ for unitaries being displacement operators as a function of $|\alpha_1|$ and $|\alpha_2|$, for $\omega=1.0$, $\theta_C=\pi/2$, $\varphi_C=0$ ($\rho_C=\ketbra{+}{+}_C$), $\theta=0$, $\phi_1-\phi_2=\pi/2$ and for different values of $|t|$. Dashed red lines in the plots show whenever $\Delta_\text{QS} = 0$, delimiting the borders of the regions where state activation is possible and impossible.}
        \label{fig:deltaQSdisplacementalpha1alpha2}
    \end{figure}

On the other hand, one has the situation after measuring the control qubit. Again, no simple analytical expression might be obtained. Nonetheless, the numerical minimization of $\Delta_{S,M}$ is done depending on the state of the control and of the measurement basis. For instance, consider the situation when the state of the control is the $\ketbra{+}{+}_C$ state ($\theta_C=\pi/2$ and $\varphi_C=0$) and the measurement state is on the $xy$-plane ($\theta_M = \pi/2$). Then, the plot of the minimum value of $\Delta_{S,M}$ for different inverse temperatures $\beta$, as a function of $\varphi_M$ is found in Fig. \ref{fig:minDeltaSMvarphiM}. It shows that when we use the basis $\{\ket{+}, \ket{-}\}$, corresponding to $\varphi_M=0$ and $\varphi_M=\pi$ respectively, to measure the control qubit, \emph{no} state activation might be achieved, for whatever values of $\lambda_k$, $\gamma_k$ and $\delta_k$. Moreover, since $\text{min}\,\Delta_{S,M}$ changes with the inverse temperature up to a constant, we take $\beta\to0$, then $\Delta_{S,M}$ becomes
\begin{equation}
    \Delta_{S,M}^{\beta\to0} = \omega \frac{f(\lambda_1,\gamma_1,\delta_1,\lambda_2,\gamma_2,\delta_2)\sin\varphi_M}{32 + g(\lambda_1,\gamma_1,\delta_1,\lambda_2,\gamma_2,\delta_2)\cos\varphi_M}
\end{equation}
with complicated functions $f(\lambda_1,\gamma_1,\delta_1,\lambda_2,\gamma_2,\delta_2)$ and $g(\lambda_1,\gamma_1,\delta_1,\lambda_2,\gamma_2,\delta_2)$. Nonetheless, we know their minimum value to be:
\begin{equation}
    \text{min}\,f(\lambda_1,\gamma_1,\delta_1,\lambda_2,\gamma_2,\delta_2) = -16
\end{equation}
and 
\begin{equation}
    \text{min}\,g(\lambda_1,\gamma_1,\delta_1,\lambda_2,\gamma_2,\delta_2) \sim -8.57.
\end{equation}
Here the $\sin\varphi_M$ on the numerator shows how $\Delta_{S,M}^{\beta\to0}=0$ when $\varphi_M = 0,\pi$.

\subsection{Quantum harmonic oscillator}
\label{sec:qho}

Continuing, now we pass to the situation in which the system is a one-mode quantum harmonic oscillator and the control is still a two-level system (qubit). The Hamiltonian of the control is the same as in the previous sections and the Hamiltonian of the system is
\begin{equation}
    H_S = \omega \left ( a^\dagger a + \frac{\mathds{1}_S}{2} \right ),
\end{equation}
where $a$ ($a^\dagger$) is the annihilation (creation) operator and as before we consider that the system and the control are resonant (same excitation energy $\omega$). The total Hamiltonian is simply the sum of the individual Hamiltonians, as no interaction is assumed between system and control. The initial state of the control is the coherent $\rho_C = \ketbra{+}{+}_C$ state ($\theta_C=\pi/2, \varphi_C=0$) and the state of the system is the thermal Gibbs state:
\begin{equation}
    \rho_S =\frac{e^{-\beta H_S}}{Z_S} 
    = (1 - e^{-\beta \omega}) \sum_n e^{-\beta \omega n} \ketbra{n}{n}_S
\end{equation}
with $Z_S = \tr \{ e^{-\beta H_S} \} = 1/(e^{\beta \omega/2} - e^{-\beta \omega/2})$ being the partition function of the system and $\ket{n}_S$ is the energy eigenstate of the system containing $n$ excitations. As usual, the initial state of system plus control is the separable state $\rho_{SC} = \rho_S \otimes \rho_C$.

When it comes to the unitaries, we might consider different cases: (i) first when both are displacement operators and (ii) second when one is a displacement operator and the other the squeeze operator.

\subsubsection{Displacement operators}
\label{sec:displacement}

We start then with both unitaries being displacement operators, that is \cite{Agarwal2012}
\begin{equation}
    U_{k} = D(\alpha_{k}) = e^{\alpha_{k}a^\dagger - \alpha_{k}^* a}, \text{for} \, \, k=1,2
\end{equation}
where $\alpha_k = |\alpha_k|e^{i \phi_k} \in \mathbb{C}$ as well as $a$ and $a^\dagger$ annihilation and creation operators, respectively. We find that
\begin{equation}
    \Delta_\text{QS} = \omega |\alpha'|^2 + |t| \big [ \text{cos}(\theta - \varphi_C + 2 |\alpha_1||\alpha_2| \sin(\phi_1 - \phi_2) ) - \cos (\theta - \varphi_C)\big]\sin\theta_C
\end{equation}
with $\alpha' = \alpha_1 + \alpha_2$ (detailed calculations can be found in Appendix \ref{app:displacementops}). This result is interesting, as it shows that whenever $\alpha_1$ and $\alpha_2$ are parallel/anti-parallel in phase-space ($\phi_1 - \phi_2 = \pi\, m$, $m \in \mathbb{Z}$), $\Delta_\text{QS}$ is always non-negative even with coherences in the control state and $|t|>0$. It reinforces the fact that these are indeed necessary, but not sufficient conditions for state activation. Also, we see that $\Delta_\text{QS}$ does not depend on the inverse temperature $\beta$, what is also something surprising. Plots of the previous equation for specific values of the parameters when $\phi_1 - \phi_2 = \pi (2m-1)/2, m \in \mathbb{Z}$ can be found in Fig. \ref{fig:deltaQSdisplacementalpha1alpha2}. These plots show that the higher $|t|$, the lower $|\alpha_1|$ and $|\alpha_2|$ are necessary for state activation.

\begin{figure}[h!]
    \centering
    \includegraphics[width=0.4\textwidth]{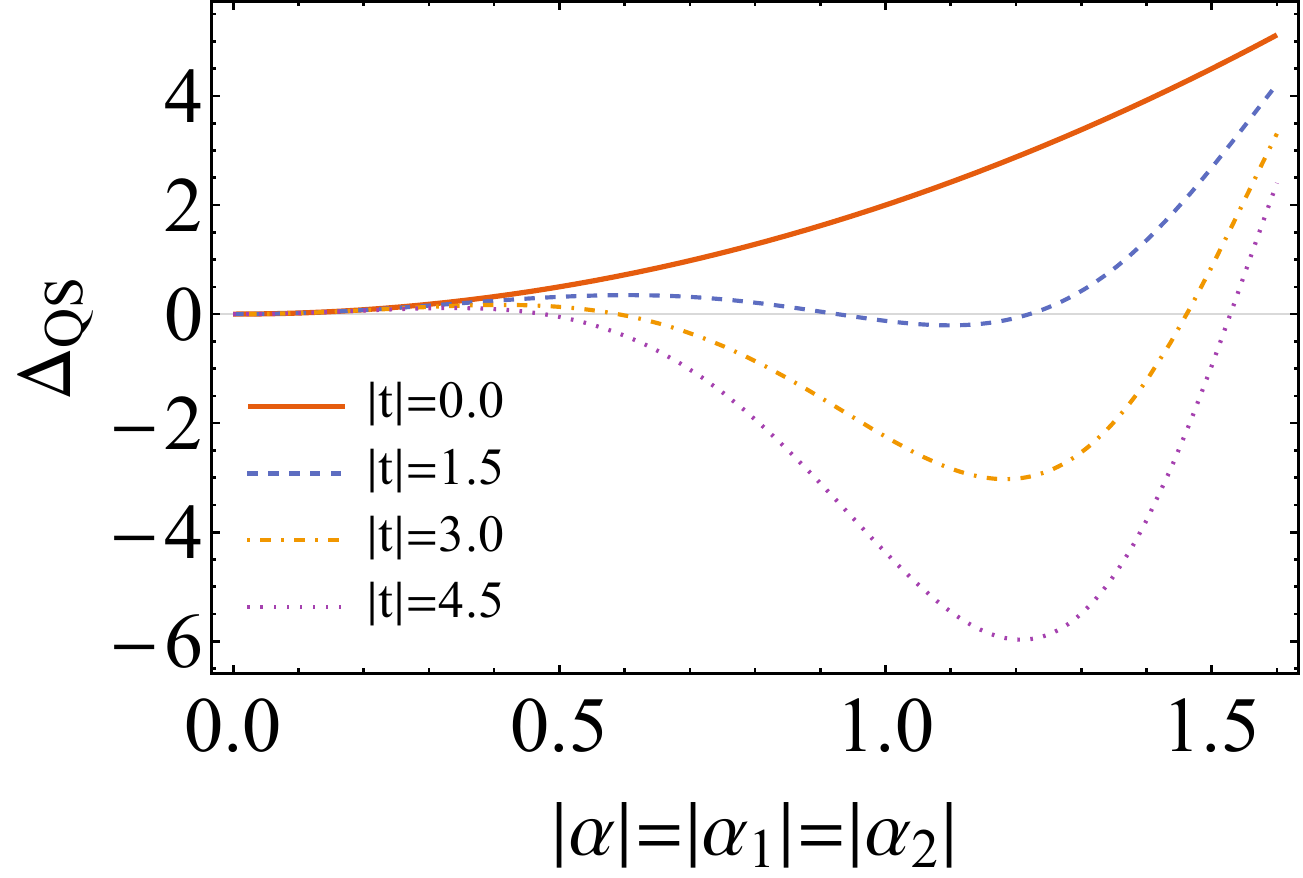}
    \caption{The plots of $\Delta_\text{QS}$ in the case that $|\alpha_1|=|\alpha_2|=|\alpha_2|$, $\omega=1.0$, $\theta=\varphi_C=0$ and $\phi_1 - \phi_2 = \theta_C = \pi/2$, and for different $|t|$. Clearly one sees that activation of the system plus control state is only possible for a range of $|\alpha|$, with a specific $|\alpha|_\text{min}$ leading to minimum $\Delta_\text{QS}$.}
    \label{fig:deltaQSdisplacementalpha}
\end{figure}

The simplifying case in which $\phi_1 - \phi_2 = \pi (2k-1)/2, k \in \mathbb{Z}$, $\theta=0$, $|\alpha_1|=|\alpha_2|=|\alpha|$ and $\theta_C=\pi/2$, $\varphi_C=0$ ($\rho_C=\ketbra{+}{+}_C$) leads to the following expression for $\Delta_\text{QS}$:
\begin{equation}
    \Delta_\text{QS} = 2\left(\omega |\alpha|^2 - |t| \sin^2\left(|\alpha|^2\right)\right).
\end{equation}
The latter is plotted in Fig. \ref{fig:deltaQSdisplacementalpha}. It shows that state activation happens for a limited range of $|\alpha|$, depending on the value of $|t|$. Moreover, it is possible to determine what is the value of $|\alpha|$ for which $\Delta_\text{QS}$ is minimum. It corresponds to:
\begin{equation}
    |\alpha|_\text{min} = \sqrt{\frac{\pi-\arcsin(\omega/|t|)}{2}}
\end{equation}
and it points to the fact that there are only solutions for $\omega \leq |t|$ and when $|t| >> \omega$, $|\alpha|_\text{min} \sim \sqrt{\pi/2}$.

Now we check whether after measuring the control in the basis $\ket{\psi_M}_C = \cos(\theta_M/2)\ket{0}_C + e^{i\varphi_M}\sin(\theta_M/2)\ket{1}_C$ the state of the system can be activated. Calculations (Appendix \ref{app:displacementops}) lead to
\begin{equation}
    \Delta_{S,M} = \omega |\alpha'|^2 \geq 0, \quad \forall \theta_C, \varphi_C, \theta_M, \varphi_M
\end{equation}
meaning that for \emph{any} control state and \emph{any} measurement state, applying displacement operators in a quantum switch setup does not activate \emph{any} passive state. This result is a consequence of the fact that displacement operators have a particular commutation relation,
\begin{equation}
    [D(\alpha_1), D(\alpha_2)] = (1-\chi)D(\alpha_1)D(\alpha_2)
\end{equation}
which shows that they almost commute, differing by the complex number $\chi$ (where $0\leq |\chi|\leq 1$). The physical implication of this mathematical property is that, when applying displacement operators in different orders to the thermal state, the final displaced state is the same in both cases up to a global complex phase. This state is clearly passive as well and no matter what post-selection is chosen, no work can be extracted from it. Previously it was not the case, because the non-diagonal element of the Hamiltonian of the control allowed the use of the coherence in the control for state activation.

\begin{figure}[t]
    \centering
    \includegraphics[width=\textwidth]{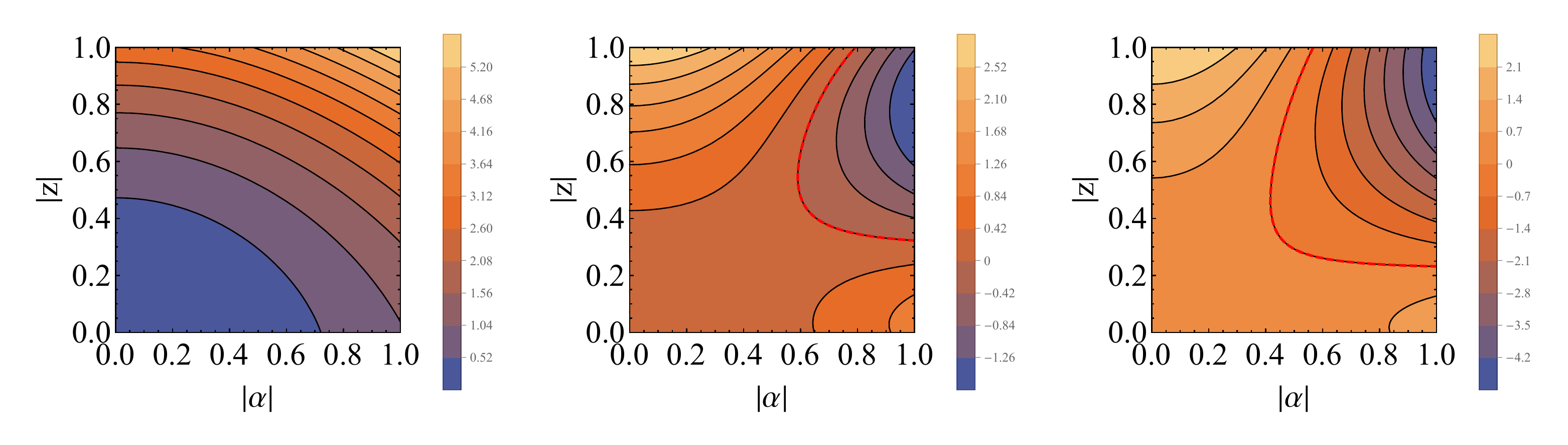}
    \caption{Plots of $\Delta_\text{QS}$ for unitaries being  the displacement and squeeze operators as a function of $|\alpha|$ and $|z|$, in the case that $\omega=\beta=1.0$, $\theta_C=\pi/2$, $\varphi_C=0$ ($\rho_C=\ketbra{+}{+}_C$), $\theta=\phi=\xi=0$ and for different values of $|t|$: (left) $|t|=0$, (center) $|t|=20$ and (right) $|t|=30$. Dashed red lines represent the situation when $\Delta_\text{QS} = 0$, delimiting the borders of the regions where state activation is possible and impossible.}
    \label{fig:deltaQSdispsqueezealphaz}
\end{figure}

\subsubsection{Displacement operator and squeeze operator}
\label{sec:displacementsqueeze}

Moreover, we might choose two different unitaries to compose our QS. One unitary is taken to be the displacement operator and the other one the squeeze operator \cite{Agarwal2012}:
\begin{equation}
    U_1 = D(\alpha) = e^{\alpha a^\dagger - \alpha^* a}
\end{equation}
\begin{equation}
    U_2 = S(z) = e^{(z a^\dagger a^\dagger - z^* a\,a)/2}
\end{equation}
where $\alpha = |\alpha|e^{i\phi}, z=|z|e^{i\xi} \in \mathbb{C}$, and again $a$ and $a^\dagger$ are annihilation and creation operators, respectively. Here the calculations are even more lengthy and as previously the details are shown in Appendix \ref{app:displacementsqueeze}. The final energy difference of system plus control is
\begin{equation}
    \Delta_{\text{QS}} = 
    \frac{\omega}{2} + \omega |\alpha|^2 + \frac{\omega \cosh(2|z|)}{2} + 2\omega\langle n \rangle_\text{th} \sinh^2|z| + \omega |\alpha|^2 \cos^2\left(\frac{\theta_C}{2}\right) \cos(\xi-2\phi)\sinh(2|z|) +  |t|\sin\theta_C\Re \{ e^{i(\varphi_C-\theta)} (\chi - 1) \}.
\end{equation}
where
\begin{equation}
\label{eq:chidisplacementsqueeze}
    \chi = \braket{\gamma}{\alpha}e^{-\langle n \rangle_\text{th} |\alpha - \gamma|^2}
\end{equation}
with $\langle n \rangle_\text{th} = 1/(e^{\beta \omega} - 1)$ being the thermal boson occupation number, $\gamma$ arises from the combination of displacement and squeeze operators
\begin{equation}
    \gamma = |\alpha| e^{i\phi} \cosh|z| - |\alpha|e^{i(\xi - \phi)}\sinh|z|
\end{equation}
and
\begin{equation}
    \braket{\gamma}{\alpha} = e^{\gamma^* \alpha - |\alpha|^2/2 - |\gamma|^2/2}
\end{equation}
The full expression is not easy to visualize, hence we show by plots how $\Delta_\text{QS}$ behaves with the parameters assuming specific values  (Fig. \ref{fig:deltaQSdispsqueezealphaz}). In comparison with the case of two displacement operators, one sees here that in order to achieve activation of the Gibbs thermal state, the value of $|t|$ must be considerably higher (one order of magnitude above $\omega$ and $\beta$). 

\begin{figure}[t]
    \centering
    \includegraphics[width=\textwidth]{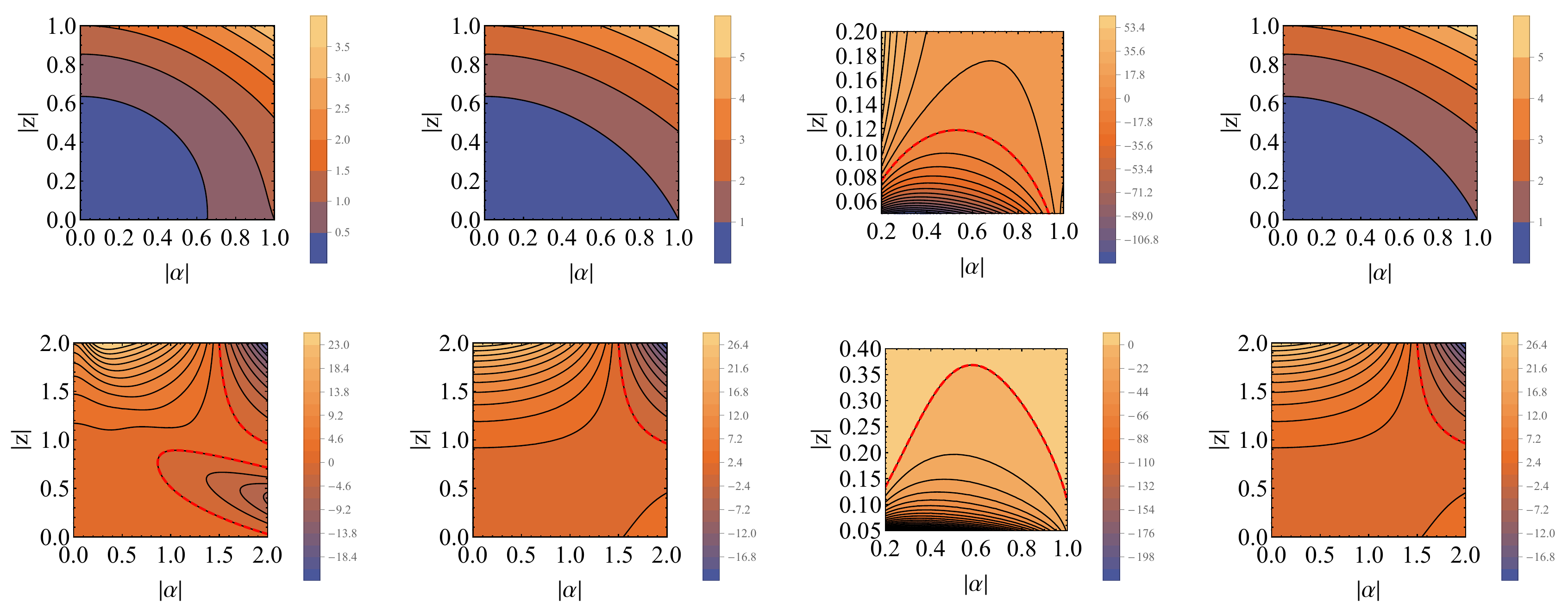}
    \caption{Plots of $\Delta_{S,M}^0$ (first row) and $\Delta_{S,M}^\pi$ (second row) as functions of $|\alpha|$ and $|z|$, for $\omega=\beta=1.0$, $\theta_C=\pi/2$, $\varphi_C=0$ ($\rho_C=\ketbra{+}{+}_C$), $\theta_M=\pi/2$ and four different $\varphi_M:$ $0,\pi/2,\pi,3\pi/2$ (each column). As in previous plots, the dashed red lines indicate when $\Delta_{S,M}^{0,\pi}=0$. Here one sees that when $\xi-2\phi=0$ (first row), the only situation when state activation is achieved is when the measurement state is the $\ketbra{-}{-}$ state ($\varphi_M = \pi$). Nonetheless, this is a delicate scenario, since for low $|\alpha|$ and $|z|$ the value of the denominator converges to zero faster than the numerator (possible divergence). On the other hand, when $\xi-2\phi=\pi$ (second row), it is always possible to activate the state of the system, where caution must also must be taken when $\varphi_M=\pi$.}
    \label{fig:deltaSMdispsqueeze}
\end{figure}

After measuring the control qubit in the state $\ket{\psi_M}_C = \cos(\theta_M/2)\ket{0}_C + e^{i\varphi_M}\sin(\theta_M/2)\ket{1}_C$, the final energy difference of the system, as shown in Appendix \ref{app:displacementsqueeze} is equal to:
\begin{equation}
    \begin{aligned}
        \Delta_{S,M} &= \frac{1}{N_M}\Bigg [ \frac{\omega}{4}(1 + \cos\theta_C \cos\theta_M)(2|\alpha|^2 + (2\langle n \rangle_\text{th} + 1)(\cosh(2|z|) - 1))\\
        &+ \omega |\alpha|^2 \cos^2\left( \frac{\theta_C}{2} \right ) \cos^2\left ( \frac{\theta_M}{2} \right ) \cos(\xi - 2\phi)\sinh(2|z|) + \frac{1}{2}\sin\theta_C \sin\theta_M \Re\{ \Delta_F e^{i(\varphi_C + \varphi_M)} \} \Bigg ]
    \end{aligned}
\end{equation}
with
\begin{equation}
    \Delta_F = \omega \chi (\gamma^*\alpha + (2\gamma^*\alpha - |\gamma|^2 - |\alpha|^2)\langle n \rangle_\text{th} - |\alpha-\gamma|^2\langle n \rangle^2_\text{th})
\end{equation}
\begin{equation}
    N_M = \frac{1}{2}(1 + \cos\theta_C \cos\theta_M + \sin\theta_C \sin\theta_M\Re \{ \chi e^{i(\varphi_C + \varphi_M)})  \}
\end{equation}
and $\chi$ the same as before (Eq. (\ref{eq:chidisplacementsqueeze})). From now on we analyze two cases: (i) $\xi-2\phi=0$:
\begin{equation}
\label{eq:deltaSM0}
    \begin{aligned}
    \Delta_{S,M}^{0} &= \frac{1}{N_M^{0}}\Bigg [ \frac{\omega}{4}(1 + \cos\theta_C \cos\theta_M)(2|\alpha|^2 + (2\langle n \rangle_\text{th} + 1)(\cosh(2|z|) - 1)) + \omega |\alpha|^2 \cos^2\left( \frac{\theta_C}{2} \right ) \cos^2\left ( \frac{\theta_M}{2} \right ) \sinh(2|z|)\\
    &- \frac{\omega|\alpha|^2}{2}\sin\theta_C \sin\theta_M e^{-2|z|-|\alpha|^2e^{-|z|}(2\langle n \rangle_\text{th}+1)(\cosh|z|-1)} \bigg (  \langle n \rangle^2_\text{th} (e^{2|z|} - 2e^{|z|} + 1) + \langle n \rangle_\text{th} (e^{2|z|} - 2e^{|z|} + |\alpha|^2 e^{-2|z|}) - e^{|z|}\bigg ) \cos(\varphi_C + \varphi_M) \Bigg ]
    \end{aligned}
\end{equation}
where
\begin{equation}
    N_M^{0} = \frac{1}{2} \Bigg(1 + \cos\theta_C \cos\theta_M + \sin\theta_C \sin\theta_M  e^{-2 |\alpha|^2 \sinh^2(|z|/2)(\cosh|z| - \sinh|z|)}
    \cos(\varphi_C + \varphi_M) \Bigg)
\end{equation}
and (ii) $\xi-2\phi=\pi$:
\begin{equation}
\label{eq:deltaSMpi}
    \begin{aligned}
    \Delta_{S,M}^{\pi} &= \frac{1}{N_M^{\pi}}\Bigg [ \frac{\omega}{4}(1 + \cos\theta_C \cos\theta_M)(2|\alpha|^2 + (2\langle n \rangle_\text{th} + 1)(\cosh(2|z|) - 1)) - \omega |\alpha|^2 \cos^2\left( \frac{\theta_C}{2} \right ) \cos^2\left ( \frac{\theta_M}{2} \right ) \sinh(2|z|)\\
    &- \frac{\omega|\alpha|^2}{2}\sin\theta_C \sin\theta_M e^{-\frac{|\alpha|^2}{2}(e^{|z|}-1)^2(2\langle n \rangle_\text{th}+1)} \bigg (  \langle n \rangle^2_\text{th} (e^{2|z|} - 2e^{|z|} + 1) + \langle n \rangle_\text{th} (|\alpha|^2e^{4|z|} - 2e^{|z|} + 1) - e^{|z|}\bigg ) \cos(\varphi_C + \varphi_M)\Bigg ]
    \end{aligned}
\end{equation}
with
\begin{equation}
    N_M^{\pi} = \frac{1}{2} \Bigg(1 + \cos\theta_C \cos\theta_M + \sin\theta_C \sin\theta_M e^{-\frac{|\alpha|^2}{2}(e^{|z|}-1)^2(2\langle n \rangle_\text{th}+1)}
    \cos(\varphi_C + \varphi_M) \Bigg).
\end{equation}
The plots of these expressions can be found in Fig. \ref{fig:deltaSMdispsqueeze}. They show that, when $\xi-2\phi=0$, one can activate the state of the system after performing the measurement on the $\ketbra{-}{-}$ state ($\varphi_M=\pi$). In this case, one must be careful, as $|z|$ and $|\alpha|$ go to zero the denominator $N_M^0$ tends to zero faster then the denominator and divergences occur. When measuring with other angles $\varphi_M$, no state activation occurs. On the other hand, for $\xi-2\phi=\pi$ it is possible to get $\Delta_{S,M}^\pi < 0$ for all $\varphi_M$ (here as before  the $\varphi_M=\pi$ case must be taken with care).

\begin{figure*}[t]
    \centering
    \includegraphics[width=0.8\textwidth]{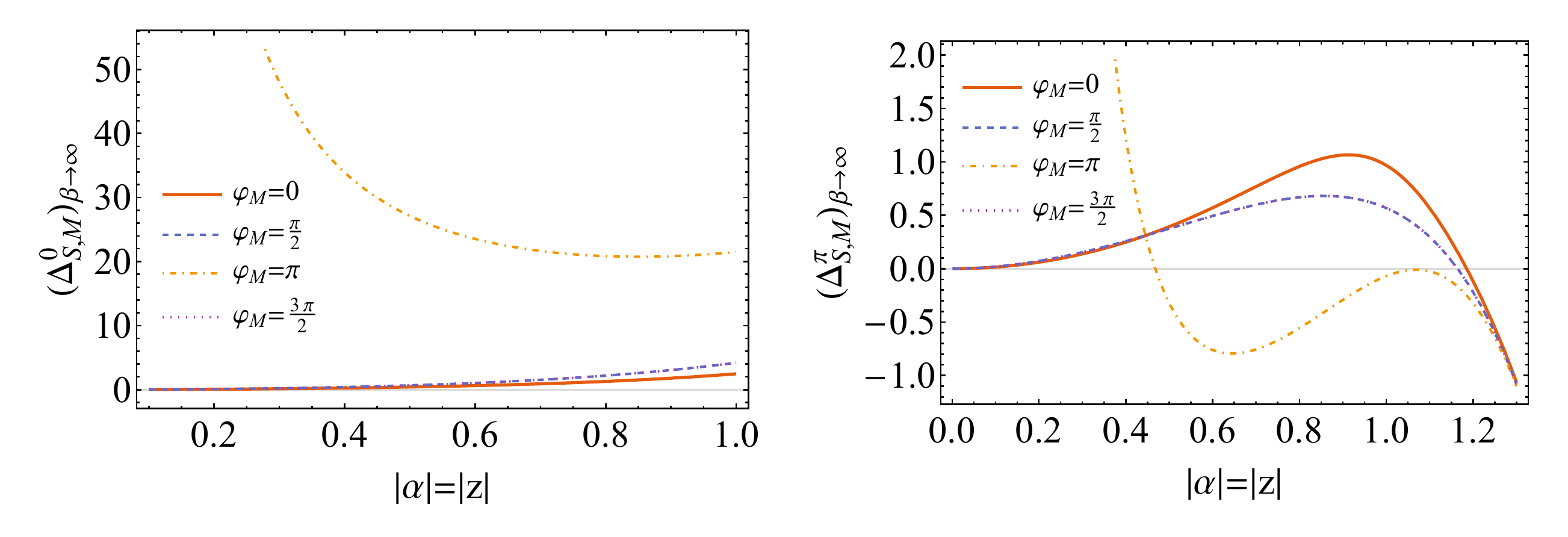}
    \caption{Plots of $\Delta_{SM}^0$ (left) and $\Delta_{SM}^\pi$ (right) when $\beta\to\infty$ as functions of $|\alpha|=|z|$, for $\omega=1.0$, $\theta_C=\pi/2$ and $\varphi_C=0$ ($\rho_C=\ketbra{+}{+}_C$), $\theta_M=\pi/2$, as well as for different values of $\varphi_M$. In this situation one sees that when $\xi-2\phi=0$ no state activation is possible, for whatever choice of measurement state. On the other hand, when $\xi-2\phi=\pi$ all the measurement choices enable state activation. For both, the cases $\varphi_M=\pi/2$ and $\varphi_M=3\pi/2$ coincide and when $|\alpha|=|z|\to 0$ there is divergence of $\Delta_{S,M}^{0,\pi}$.}
    \label{fig:deltaSMdispsqueezebetainf}
\end{figure*}

As a last scenario to be evaluated, consider when $\beta\to\infty$, which corresponds to:
\begin{equation}
    \langle n \rangle_\text{th} \to 0
\end{equation}
and the Gibbs thermal state tends asymptotically to the ground state $\rho_S\to \ketbra{0}{0}_S$. Then, Eqs. (\ref{eq:deltaSM0}) and (\ref{eq:deltaSMpi}) are simplified accordingly. The final expressions are plotted in Fig. \ref{fig:deltaSMdispsqueezebetainf} in the case that the magnitude of the displacement and the squeeze are the same $|\alpha|=|z|$, $\rho_C = \ketbra{+}{+}_C$ and $\theta_M=\pi/2$. The plots show that in this limit, it is impossible to achieve state activation for any chosen measurement state when $\xi-2\phi=0$. Nevertheless, in the situation that $\xi-2\phi=\pi$ one can get negative values of $\Delta_{S,M}^\pi$ for any of the chosen measurement angles $\varphi_M$. Here, as before, when $|\alpha|=|z|\to 0$ and one measures in the $\ketbra{-}{-}$ state ($\varphi_M=\pi$), the values of $\Delta_{S,M}^{0,\pi}$ diverge. Finally, it is noticeable that $\varphi_M=\pi/2$ and $\varphi_M=3\pi/2$ lead to the same results.

\section{Conclusions}
\label{sec:conclusions}

In this paper we focus on the problem of quantifying state activation in scenarios where the quantum switch (QS) is applied. It is well known that passive state in quantum thermodynamics cannot be activated by any unitary operations~\cite{Haag1974,Lenard1978,Pusz1978,Allahverdyan2004,Safranek2023}, needing extra resources (e.g. coherences in the state) to be able to be used in thermodynamical tasks. With the rapid advancement of indefinite causal order (ICO) research in its quantum switch (QS) form~\cite{Araujo2014,Quintino2019,Renner2022,Guerin2016,Ebler2018,Wei2019,Zhao2020,ChapeauBlondeau2021,Xie2021,ChapeauBlondeau2022,Yin2023,Chiribella2021,Abbott2020,Guerin2019,Jia2019,Liu2023,ChapeauBlondeau2021b,Procopio2015,Rubino2017,Araujo2015,Goswami2018,Goswami2020,Guo2020,Rubino2021,Taddei2021,Caleffi2020}, especially in the context of communication, computation and metrology, the matter of its resourcefulness in performing thermodynamic tasks is still open, apart from a few works on this topic~\cite{Felce2020,Guha2020,Simonov2022,Dieguez2023}. We then put forward the result that, the QS by itself \emph{does not} ensure state activation, needing resources outside of itself to enable energy extraction from passive states (Sec. \ref{sec:results}). In order to activate the composite state of system plus control, non-diagonal elements (in the computational basis defined by the action of QS) on both the state of the control and the Hamiltonian of the control are necessary (but not sufficient) conditions. The first corresponds to \emph{coherences} and the second is related to the inner transitions between the two states of the control. Moreover, when the latter does not exist, it is still possible to activate the state of the system alone by measuring the state of the control. The measurement that ensures energy extraction is case dependent (what is the kind of system considered, what are the unitaries, etc.). In examples (Sec. \ref{sec:examples}) we then present different scenarios where state activation happens or does not happen, always taking as a reference point for the state of the system the Gibbs thermal state. These results point to the fact that, when considering thermodynamical tasks involving the QS, one must always consider a very specific setup, where all parameters are well controlled, otherwise, it is difficult to predict whether the QS is in fact bringing anything new to what is already known in quantum thermodynamics.

This work points to a few future possible studies. First, it is still necessary to quantify the energetic cost of measuring the control qubit. From \emph{Landauer's principle} we know that to every measurement corresponds some finite dissipation of heat. Thus, it is imperative to compare this energetic cost to what is gained from state activation by means of measurement of the control qubit. This comparison will present what are the situations where a net energetic gain is obtained. Second, the QS is not the only example of indefinite causal order. Using process matrices~\cite{Oreshkov2012}, one is able to devise situations where causal separability does not hold and causal inequalities are violated~\cite{Araujo2015}. Still, little is known about these scenarios and what are their possible use cases. An interesting open problem for the near future is thus to understand how they act on passive states and whether they are capable of activating such states without the need of extra resources.

\section*{Acknowledgements} 
We acknowledge support by the Foundation
for Polish Science (IRAP project, ICTQT, contract no.
2018/MAB/5, co-financed by EU within Smart Growth
Operational Programme).

\bibliography{References}

\appendix
{\color{black}
\section{\label{app:displacementops}Displacement operators}

The joint energy variation of system plus control after applying the quantum switch of two displacement operators is equal to ($\bra{1} H_C \ket{0} = t^*$, $\bra{0} \rho_C \ket{1} = e^{i\varphi_C}\sin(\theta_C/2)\cos(\theta_C/2)$):
\begin{eqnarray}
    \Delta_\text{QS} &=& \Delta_S + 2 \Re \{ \bra{0} \rho_C \ketbra{1}{1} H_C \ket{0} (\chi - 1) \} \\
    &=& \Delta_S + \sin\theta_C \Re \{e^{i \varphi_C} t^* (\chi - 1) \}
\end{eqnarray}
where
\begin{equation}
    \Delta_S = \cos^2\left(\frac{\theta_C}{2}\right) E_{12} + \sin^2\left(\frac{\theta_C}{2}\right) E_{21} - E_S
\end{equation}
with 
\begin{equation}
    E_{12} = \tr \{ D(\alpha_2)D(\alpha_1)\rho_S D^\dagger(\alpha_1)D^\dagger(\alpha_2)H_S \},
\end{equation}
\begin{equation}
    E_{21} = \tr \{ D(\alpha_1)D(\alpha_2)\rho_S D^\dagger(\alpha_2)D^\dagger(\alpha_1)H_S \},
\end{equation}
\begin{eqnarray}
    E_S &=& \tr \{ \rho_S H_S \} \\
    &=& \omega (1 - e^{-\beta \omega}) \sum_n e^{-\beta \omega n} \bra{n} \left ( a^\dagger a + \frac{\mathds{1}}{2} \right ) \ket{n}\\
    &=& \omega \left ( \langle n \rangle_\text{th} + \frac{1}{2}\right ).
\end{eqnarray}
where
\begin{equation*}
    \langle n \rangle_\text{th} = \frac{1}{e^{\beta \omega} - 1}
\end{equation*}
is the thermal boson occupation number (or \emph{Bose-Einstein distribution} with zero chemical potential) and $\chi$ is connected to the cross-map unitary (check main text for definition). Moreover,
\begin{eqnarray}
    E_{21} &=& \tr \{ D(\alpha_1)D(\alpha_2)\rho_S D^\dagger(\alpha_2)D^\dagger(\alpha_1)H_S \}\\
    &=& \frac{\omega}{2} + \omega \tr \{ \rho_S D^\dagger (\alpha_2) D^\dagger (\alpha_1)  a^\dagger a  D(\alpha_1) D(\alpha_2) \} \\
    &=& E_S + \omega |\alpha'|^2
\end{eqnarray}
where $\alpha':=\alpha_1 + \alpha_2$ and we used the fact that $D^\dagger (\alpha) a^\dagger D(\alpha) = a^\dagger + \alpha^*$, $D^\dagger (\alpha) a D(\alpha) = a + \alpha$ \cite{Agarwal2012}. Similarly,
\begin{eqnarray}
    E_{12} &=& \tr \{ D(\alpha_2)D(\alpha_1)\rho_S D^\dagger(\alpha_1)D^\dagger(\alpha_2)H_S \} \\
    &=& E_S + \omega |\alpha'|^2.
\end{eqnarray}
Moreover,
\begin{eqnarray}
    \chi &=& \tr \{ D(\alpha_2)D(\alpha_1)\rho_S D^\dagger (\alpha_2)D^\dagger(\alpha_1) \}\\
    &=& \tr \{ e^{\alpha_1^* \alpha_2 - \alpha_1 \alpha_2^*} D(\alpha_1)D(\alpha_2)\rho_S D^\dagger(\alpha_2)D^\dagger(\alpha_1) \}\\
    &=& e^{\alpha_1^* \alpha_2 - \alpha_1 \alpha_2^*}
\end{eqnarray}
in which we use the following relation \cite{Agarwal2012}:
\begin{equation*}
    D(\alpha_2)D(\alpha_1) = e^{\alpha_1^* \alpha_2 - \alpha_1 \alpha_2^*}D(\alpha_1)D(\alpha_2).
\end{equation*}
Hence
\begin{equation}
    \Delta_\text{QS} = \omega |\alpha'|^2 + |t| (\text{cos}(\theta - \varphi_C + 2 |\alpha_1||\alpha_2| \sin(\phi_1 - \phi_2) ) - \cos (\theta-\varphi_C))\sin\theta_C
\end{equation}

After measuring the control qubit, the final state of the system contains coherence terms which contribute to the final energy of the system. The final energy difference of the system post-measurement is written as
\begin{equation}
        \Delta_{S,M} = \frac{1}{N_M} \bigg ( \cos^2\left(\frac{\theta_C}{2}\right)\cos^2\left(\frac{\theta_M}{2}\right) \Delta_{12} + \sin^2\left(\frac{\theta_C}{2}\right)\sin^2\left(\frac{\theta_M}{2}\right) \Delta_{21} + \frac{1}{2} \sin\theta_C \sin\theta_M \Re \{ \Delta_F e^{i(\varphi_C + \varphi_M)} \} \bigg )
\end{equation}
where 
\begin{equation}
    N_M = \frac{1}{2}(1 + \cos\theta_C \cos\theta_M + \sin\theta_C \sin\theta_M\Re \{ \chi e^{i(\varphi_C + \varphi_M)})  \}
\end{equation}
is a normalization constant and
\begin{equation}
    \Delta_{12} := \tr \{ D(\alpha_2) D(\alpha_1) \rho_S D^\dagger(\alpha_1) D^\dagger(\alpha_2) \} - E_S
\end{equation}
\begin{equation}
    \Delta_{21} := \tr \{ D(\alpha_1) D(\alpha_2) \rho_S D^\dagger(\alpha_2) D^\dagger(\alpha_1) \} - E_S
\end{equation}
\begin{equation}
    \Delta_F := F_S - \chi E_S
\end{equation}
being $F_S$ defined by
\begin{equation}
    F_S := \tr \{D(\alpha_2) D(\alpha_1) \rho_S D^\dagger(\alpha_2) D^\dagger(\alpha_1) H_S\}.
\end{equation}
As a first step to calculate $\Delta_{S,M}$, we find that
\begin{eqnarray}
    F_{S} &=& \tr \{ D(\alpha_2) D(\alpha_1) \rho_S D^\dagger (\alpha_2)D^\dagger(\alpha_1)H_S \}\\
    &=& \chi \tr\{D(\alpha_1) D(\alpha_2) \rho_S D^\dagger(\alpha_2)D^\dagger(\alpha_1)H_S \} \\
    &=& \chi E_{21}\\
    &=& \chi (E_S + \omega |\alpha'|^2)
\end{eqnarray}
and then 
\begin{equation}
    \Delta_F = \chi\omega|\alpha'|^2
\end{equation}
The values of $E_{12}$ and $E_{21}$ are the same as before, hence
\begin{equation}
    \Delta_{12}=\Delta_{21}=\omega|\alpha'|^2.
\end{equation}
Finally, putting everything together:
\begin{equation}
    \Delta_{S,M} = \omega |\alpha'|^2 \geq 0
\end{equation}
and then it is impossible to activate the state of the system by measuring the state of the control, for whatever choice of measurement state. This result is a consequence of the fact that two displacement operators have a very specific commutation relation,
\begin{equation}
    [D(\alpha_1), D(\alpha_2)] = (1-e^{\alpha_1^* \alpha_2 - \alpha_1 \alpha_2^*})D(\alpha_1)D(\alpha_2).
\end{equation}
The physical implication of this property is discussed in the main text.

{\color{black}
\section{\label{app:displacementsqueeze}Displacement operator and squeeze operator}

In the case that one of the unitaries being the displacement operator and the squeeze operator the final energy difference is
\begin{equation}
    \Delta_\text{QS} = \Delta_S + \sin\theta_C\Re \{ e^{i\varphi_C} t^* (\chi - 1) \}
\end{equation}
where 
\begin{equation}
    \Delta_S = \cos^2\left(\frac{\theta_C}{2}\right) E_{12} + \sin^2\left(\frac{\theta_C}{2}\right) E_{21} - E_S
\end{equation}
and
\begin{equation}
    E_{12} = \tr \{ S(z) D(\alpha) \rho_S D^\dagger (\alpha) S^\dagger (z) H_S \},
\end{equation}
\begin{equation}
    E_{21} = \tr \{ D(\alpha) S(z) \rho_S S^\dagger (z) D^\dagger (\alpha)  H_S \},
\end{equation}
\begin{equation}
    \chi = \tr \{ S(z) D(\alpha) \rho_S S^\dagger (z) D^\dagger (\alpha) \}.
\end{equation}
Applying the relations $D^\dagger(\alpha)a^\dagger D(\alpha) = a^\dagger +\alpha^*$, $D^\dagger(\alpha)a D(\alpha) = a + \alpha$, $S^\dagger (z) a^\dagger S(z) = a^\dagger \cosh |z| + a e^{-i \xi} \sinh|z|$ and $S^\dagger (z) a S(z) = a \cosh |z| + a^\dagger e^{i \xi} \sinh|z|$ \cite{Agarwal2012}, one finds that
\begin{equation}
    E_{21} = \omega |\alpha|^2 + \frac{\omega}{2} \left (2 \langle n \rangle_\text{th} + 1 \right )\cosh(2|z|)
\end{equation}
\begin{equation}
    E_{12} = E_{21} + \omega |\alpha|^2 \cos (\xi - 2\phi)\sinh(2|z|). 
\end{equation}
Therefore,
\begin{equation}
        \Delta_S = -\frac{\omega}{2} + \omega |\alpha|^2 + \frac{\omega \cosh(2|z|)}{2} + 2\omega\langle n \rangle_\text{th} \sinh^2|z| + \omega |\alpha|^2 \cos^2\left(\frac{\theta_C}{2}\right) \cos(\xi-2\phi)\sinh(2|z|).
\end{equation}

In order to calculate $\chi$, we use the braiding relation $D(\alpha)S(z) = S(z)D(\gamma)$, where $\gamma = |\alpha| e^{i\phi} \cosh|z| - |\alpha|e^{i(\xi - \phi)}\sinh|z|$ \cite{Nieto1997}. By applying the \emph{P-representation} of the thermal state \cite{Agarwal2012},
\begin{equation*}
    \rho_S = \int P_T (\eta) \ketbra{\eta}{\eta} \dd ^2 \eta
\end{equation*}
where $\ket{\eta}$ is a coherent state and
\begin{equation}
    P_T (\eta) = \frac{1}{\pi \langle n \rangle_\text{th}} e^{-|\eta|^2/\langle n \rangle_\text{th}}
\end{equation}
then
\begin{equation}
    \chi = \int P_T(\eta) \bra{\eta} D^\dagger(\gamma) D(\alpha) \ket{\eta} \dd ^2 \eta = \int P_T(\eta) e^{\frac{\eta}{2}(\gamma^*  - \alpha^*)} e^{\frac{\eta^*}{2} (\alpha - \gamma)} \braket{\eta+\gamma}{\eta+\alpha}\dd^2 \eta 
\end{equation}
hence, by evaluating the integral:
\begin{equation}
    \chi = \braket{\gamma}{\alpha}e^{-\langle n \rangle_\text{th} |\alpha - \gamma|^2}
\end{equation}
in which we used the fact that for two coherent states \cite{Agarwal2012}:
\begin{equation*}
    \braket{\alpha}{\beta} = e^{\alpha^* \beta - |\alpha|^2/2 - |\beta|^2/2}.
\end{equation*}

Finally:
\begin{equation}
    \Delta_{\text{QS}} = 
    \frac{\omega}{2} + \omega |\alpha|^2 + \frac{\omega \cosh(2|z|)}{2} + 2\omega\langle n \rangle_\text{th} \sinh^2|z| + \omega |\alpha|^2 \cos^2\left(\frac{\theta_C}{2}\right) \cos(\xi-2\phi)\sinh(2|z|) +  |t|\sin\theta_C\Re \{ e^{i(\varphi_C-\theta)} (\chi - 1) \}.
\end{equation}
Once again, after measuring the control qubit,
the final state of the system has extra terms
add to the final energy of the system. The
final energy of the system post-measurement is written
as
\begin{equation}
        \Delta_{S,M} = \frac{1}{N_M} \bigg ( \cos^2\left(\frac{\theta_C}{2}\right)\cos^2\left(\frac{\theta_M}{2}\right) \Delta_{12} + \sin^2\left(\frac{\theta_C}{2}\right)\sin^2\left(\frac{\theta_M}{2}\right) \Delta_{21} + \frac{1}{2} \sin\theta_C \sin\theta_M \Re \{ \Delta_F e^{i(\varphi_C + \varphi_M)} \} \bigg )
\end{equation}
where
\begin{equation}
    N_M = \frac{1}{2}(1 + \cos\theta_C \cos\theta_M + \sin\theta_C \sin\theta_M\Re \{ \chi e^{i(\varphi_C + \varphi_M)})  \}
\end{equation}
\begin{equation}
    \Delta_{12} := \tr \{ S(z) D(\alpha) \rho_S D^\dagger(\alpha) S^\dagger(z) \} - E_S
\end{equation}
\begin{equation}
    \Delta_{21} := \tr \{ D(\alpha) S(z) \rho_S S^\dagger (z) D^\dagger(\alpha) \} - E_S
\end{equation}
and 
\begin{equation}
    \Delta_F := F_S - \chi E_S
\end{equation}
being $F_S$ defined by
\begin{eqnarray}
    F_S &:=& \tr \{S(z) D(\alpha) \rho_S S^\dagger (z)D^\dagger(\alpha) H_S\}\\
    &=& \omega \frac{\chi}{2} + \tr \{S(z) D(\alpha) \rho_S S^\dagger (z)D^\dagger(\alpha) a^\dagger a\} 
\end{eqnarray}
which, by using the previous braiding relation is equal to
\begin{equation}
    F_S = \omega \frac{\chi}{2} + \int P_T(\eta) \bra{\eta}D^\dagger(\gamma)a^\dagger a D(\alpha) \ket{\eta}\dd^2 \eta
\end{equation}
with 
\begin{equation*}
    \bra{\eta}D^\dagger(\gamma)a^\dagger a D(\alpha) \ket{\eta} = e^{\frac{\eta}{2}(\gamma^*  - \alpha^*)} e^{\frac{\eta^*}{2} (\alpha - \gamma)} (\eta^*+\gamma^*)(\eta+\alpha)\braket{\eta+\gamma}{\eta+\alpha}
\end{equation*}
since coherent states are eigenvectors of the annihilation operator,
\begin{equation*}
    a\ket{\alpha} = \alpha\ket{\alpha}.
\end{equation*}
Then, after computing the integral in the complex plane, one has
\begin{equation}
    F_S = \omega \chi \Bigg ( \frac{1}{2} + \gamma^* \alpha + (1+ 2\gamma^*\alpha - |\alpha|^2 - |\gamma|^2)\langle n \rangle_\text{th} - |\alpha - \gamma|^2 \langle n \rangle^2_\text{th} \Bigg)
\end{equation}
and
\begin{equation}
    \Delta_F = \omega \chi (\gamma^*\alpha + (2\gamma^*\alpha - |\gamma|^2 - |\alpha|^2)\langle n \rangle_\text{th} - |\alpha-\gamma|^2\langle n \rangle^2_\text{th}).
\end{equation}
Moreover, using the values of $E_{12}$ and $E_{21}$, one obtains
\begin{equation}
    \Delta_{21} = \omega |\alpha|^2 + \omega \langle n \rangle_\text{th} (\cosh(2|z|) - 1) + \omega \sinh^2|z|
\end{equation}
\begin{equation}
    \Delta_{12} = \Delta_{21} + \omega |\alpha|^2 \cos(\xi - 2\phi)\sinh(2|z|)
\end{equation}
and finally
\begin{equation}
    \begin{aligned}
        \Delta_{S,M} &= \frac{1}{N_M}\Bigg [ \frac{\omega}{4}(1 + \cos\theta_C \cos\theta_M)(2|\alpha|^2 + (2\langle n \rangle_\text{th} + 1)(\cosh(2|z|) - 1))\\
        &+ \omega |\alpha|^2 \cos^2\left( \frac{\theta_C}{2} \right ) \cos^2\left ( \frac{\theta_M}{2} \right ) \cos(\xi - 2\phi)\sinh(2|z|) + \frac{1}{2}\sin\theta_C \sin\theta_M \Re\{ \Delta_F e^{i(\varphi_C + \varphi_M)} \} \Bigg ]
    \end{aligned}
\end{equation}
which taking the cases (i) $\xi-2\phi=0$:
\begin{equation}
    \begin{aligned}
    \Delta_{S,M}^{0} &= \frac{1}{N_M^{0}}\Bigg [ \frac{\omega}{4}(1 + \cos\theta_C \cos\theta_M)(2|\alpha|^2 + (2\langle n \rangle_\text{th} + 1)(\cosh(2|z|) - 1)) + \omega |\alpha|^2 \cos^2\left( \frac{\theta_C}{2} \right ) \cos^2\left ( \frac{\theta_M}{2} \right ) \sinh(2|z|)\\
    &- \frac{\omega|\alpha|^2}{2}\sin\theta_C \sin\theta_M e^{-2|z|-|\alpha|^2e^{-|z|}(2\langle n \rangle_\text{th}+1)(\cosh|z|-1)} \bigg (  \langle n \rangle^2_\text{th} (e^{2|z|} - 2e^{|z|} + 1) + \langle n \rangle_\text{th} (e^{2|z|} - 2e^{|z|} + |\alpha|^2 e^{-2|z|}) - e^{|z|}\bigg ) \cos(\varphi_C + \varphi_M) \Bigg ]
    \end{aligned}
\end{equation}
where
\begin{equation}
    N_M^{0} = \frac{1}{2} \Bigg(1 + \cos\theta_C \cos\theta_M + \sin\theta_C \sin\theta_M  e^{-2 |\alpha|^2 \sinh^2(|z|/2)(\cosh|z| - \sinh|z|)}
    \cos(\varphi_C + \varphi_M) \Bigg)
\end{equation}
and (ii) $\xi-2\phi=\pi$:
\begin{equation}
    \begin{aligned}
    \Delta_{S,M}^{\pi} &= \frac{1}{N_M^{\pi}}\Bigg [ \frac{\omega}{4}(1 + \cos\theta_C \cos\theta_M)(2|\alpha|^2 + (2\langle n \rangle_\text{th} + 1)(\cosh(2|z|) - 1)) - \omega |\alpha|^2 \cos^2\left( \frac{\theta_C}{2} \right ) \cos^2\left ( \frac{\theta_M}{2} \right ) \sinh(2|z|)\\
    &- \frac{\omega|\alpha|^2}{2}\sin\theta_C \sin\theta_M e^{-\frac{|\alpha|^2}{2}(e^{|z|}-1)^2(2\langle n \rangle_\text{th}+1)} \bigg (  \langle n \rangle^2_\text{th} (e^{2|z|} - 2e^{|z|} + 1) + \langle n \rangle_\text{th} (|\alpha|^2e^{4|z|}- 2e^{|z|} + 1) - e^{|z|}\bigg ) \cos(\varphi_C + \varphi_M)\Bigg ]
    \end{aligned}
\end{equation}
with
\begin{equation}
    N_M^{\pi} = \frac{1}{2} \Bigg(1 + \cos\theta_C \cos\theta_M + \sin\theta_C \sin\theta_M  e^{-\frac{|\alpha|^2}{2}(e^{|z|}-1)^2(2\langle n \rangle_\text{th}+1)}
    \cos(\varphi_C + \varphi_M) \Bigg).
\end{equation}
\end{document}